%% file: main.tex
\documentclass[10pt,
 reprint,
 prc,
 amsmath,amssymb,
 aps
]{revtex4-2}

\usepackage{dcolumn}
\usepackage{graphicx}
\usepackage{bm}
\usepackage[subrefformat=parens]{subcaption}
\usepackage[compat=1.1.0]{tikz-feynhand}
\usepackage{physics}
\usepackage{bxtexlogo}
\usepackage[samesize]{cancel}
\usepackage{xcolor}
\usepackage{lineno}

\begin{document}

\title{$K^- d \rightarrow \pi \Lambda N$ reaction with in-flight kaons for studying the $\Lambda N$ interaction}

\author{Shunsuke Yasunaga}
\affiliation{Department of Physics, Tokyo Institute of Technology, 2-12-1 Ookayama, Meguro-ku, Tokyo 152-8551, Japan}
\author{Daisuke Jido}
\affiliation{Department of Physics, Tokyo Institute of Technology, 2-12-1 Ookayama, Meguro-ku, Tokyo 152-8551, Japan}
\author{Takatsugu Ishikawa}
\affiliation{Research Center for Nuclear Physics(RCNP), Osaka University, 10-1 Mihogaoka, Ibaraki, Osaka, 567-0047, Japan}
\date{\today}

\begin{abstract}
The $\Lambda N$ invariant mass spectra for the reactions $K^-d\rightarrow\pi^-\Lambda p$ and $K^-d\rightarrow\pi^0\Lambda n$ are calculated for experimental study of isospin symmetry breaking in the $\Lambda N$ scattering at low energies, the difference in the scattering lengths and effective ranges of $\Lambda p$ and $\Lambda n$ systems.
The calculations are performed for in-flight kaons with a momentum of 1000 MeV/c with employing partial wave analysis up to the p-wave for meson-baryon amplitudes and the spin-flip term for baryon-baryon amplitudes. Kinematic selection is utilized to suppress the background processes by selecting forward-emitting pions and higher momentum nucleons. It is worth noting that isospin symmetry breaking in the $\Lambda N$ system can be extracted from the difference of the $\Lambda N$ invariant mass spectra between the $K^-d\rightarrow\pi^-\Lambda p$ and $K^-d\rightarrow\pi^0\Lambda n$ reactions.
\end{abstract}

\maketitle

\input introduction.tex
\input formulation.tex
\input results.tex
\input discussion.tex
\input summary.tex

\section*{Acknowledgement}
We would express our gratitude Yutaro Iizawa for our contributions in early stage. This work of S.Y. was partly supported by the Advanced Research Center for Quantum Physics and Nanoscience, Tokyo Institute of Technology. The work of D.J. was partly supported by Grants-in-Aid for Scientific Research from JSPS (21K03530, 22H04917, 23K03427). The work of T.I. was partly supported by Grants-in-Aid for Scientific Research from JSPS (21H00114, 22H00124).
\input appendix.tex

\bibliographystyle{apsrev4-2}
\bibliography{Reference}

\end{document}

%% file: introduction.tex
\section{Introduction}\label{sec:int}
Isospin symmetry breaking in the $\Lambda N$ system, the difference between $\Lambda p$ and $\Lambda n$ interactions, is one of the important topics in hadronic physics. Low-energy scattering parameters, scattering lengths and effective ranges, for the $\Lambda p$ system have been extracted through the experimental data of the $\Lambda p$ final state interaction in the $pp \rightarrow K^+\Lambda p$ reaction~\cite{budzanowski_2010,hauenstein_2017}. Additionally, the $\Lambda p$ elastic scattering cross section in the high momentum region has been measured in Ref.~\cite{rowley_2021}, and the $\Lambda p$ correlation function has been studied via femtoscopy~\cite{acharya_2019}.
Whereas the equivalent low-energy parameters for the $\Lambda n$ system, however, have not been measured experimentally yet. Instead, recent experimental analyses for the $A=4$ hypernuclei, $^4_\Lambda$H and $^4_\Lambda$He, have suggested a possibility of large isospin symmetry breaking in the $\Lambda N$ system through the difference of their binding energies, $\Delta B^4_\Lambda$ = $B(^4_\Lambda{\rm He}) - B
(^4_\Lambda\rm{H})$~\cite{a1collaboration_2015,j-parce13collaboration_2015,abdallah_2022}. One theoretical work~\cite{haidenbauer_2021} performed calculations to extract values for the scattering lengths and effective ranges of the $\Lambda n$ system from $\Delta B^4_\Lambda$. An alternative approach to investigate isospin symmetry breaking in the $\Lambda N$ system is to compare the $\Lambda N$ invariant mass spectra of the $K^-d\rightarrow\pi^-\Lambda p$ reaction and the $K^-d\rightarrow\pi^0\Lambda n$ reaction~\cite{iizawa_2022}. The differences of the spectra indicate isospin symmetry breaking in the parameters of the $\Lambda N$ interaction.

There have been various studies on the $K^-d\rightarrow\pi^-\Lambda p$ reaction. Theoretical researches in the early stages were carried out in Refs.~\cite{karplus_1959,kotani_1959}. Later studies utilized experimental data from the $\bar{K}N\rightarrow\pi Y$ reaction \cite{dalitz_1981,dalitz_1982,torres_1986} and employed Faddeev formalism \cite{toker_1979,toker_1981}. Experimental observation and analyses in the case of kaons at rest were presented in Refs.~\cite{dahl_1961,tan_1969}. In the case of in-flight kaons were presented in Refs.~\cite{cline_1968,alexander_1969,eastwood_1971,braun_1977,pigot_1985}, and upcoming experiment, J-PARC E90 \cite{ichikawa_2022}. These studies mainly focused on the $\Sigma N$ cusp structure observed near the mass threshold of the $\Sigma N$ system in the $\Lambda N$ invariant mass spectra of this reaction.
In relation to the $K^-d\rightarrow\pi^-\Lambda p$ reaction, the $K^-d\rightarrow\pi\Sigma n$ reaction also has been studied. This reaction is valuable for analyzing the properties of the $\Lambda(1405)$ resonance induced by the $\bar{K}N$ channel~\cite{jido_2009}. Theoretical investigations were carried out in Refs.~\cite{jido_2009,jido_2011,jido_2013,yamagata-sekihara_2013,miyagawa_2012,miyagawa_2018,ohnishi_2016}. These calculations utilized $\bar{K}N\rightarrow\pi\Sigma$ amplitudes obtained from chiral unitary approaches~\cite{kaiser_1995,oset_1998,oller_2000,oset_2002,jido_2002,lutz_2002,jido_2003,hyodo_2003,hyodo_2004,hyodo_2012,mai_2014}. Recently, the corresponding experimental results were provided in Ref.~\cite{aikawa_2023}.

Regarding the $\Lambda N$ interaction itself, several theoretical approaches have been performed. Phenomenological approaches based on meson-exchange models have been studied in works such as Refs.~\cite{rijken_1999,rijken_2010,nagels_2019,holzenkamp_1989,reuber_1994,tominaga_1998,tominaga_2001}. Quark models have also been used to study the $\Lambda N$ system, as seen in Ref.~\cite{garcilazo_2007}, with utilizing spin-flavor SU(6) symmetry in their analysis shown in Refs.~\cite{fujiwara_1996,kohno_2000,fujiwara_2007}.
Another kind of approaches are based on effective field theory with chiral SU(3) symmetry \cite{li_2016,li_2018,song_2018,polinder_2006,haidenbauer_2013,haidenbauer_2020,haidenbauer_2021,korpa_2001,petschauer_2020,savage_1996,ren_2020}. Due to the experimental difficulties on the direct $\Lambda N$ scattering, three-body scattering such as $K^-d\rightarrow\pi\Lambda N$ reaction is a candidate to investigate the $\Lambda N$ interaction~\cite{iizawa_2022}.

In this work, we present a possibility to utilize the $K^-d\rightarrow\pi\Lambda N$ reaction to study isospin symmetry breaking in the $\Lambda N$ system phenomenologically with the low-enegy effective range expansion. This is an extension of the previous research in the case of stopped kaons~\cite{iizawa_2022}. We calculate the $\Lambda N$ invariant mass spectra of these reactions in the case of in-flight kaons with momenta 1000 MeV/c, which is accessible at J-PARC. Since the $K^-d\rightarrow\pi\Lambda N$ reactions involve the $\Lambda N$ interaction in their final states, by taking the ratio of the $\Lambda N$ invariant mass spectra of $K^-d\rightarrow\pi^-\Lambda p$ and $\pi^0\Lambda n$, we explore the sensitivity of the reaction to the rarios of scattering lengths $a_{\Lambda n}/a_{\Lambda p}$ and effective ranges $r_{\Lambda n}/r_{\Lambda p}$, respectively. For the in-flight case, $\Lambda N$ interaction is expected to contribute both of the spin-singlet and the spin-triplet terms, however, the spin-singlet term is negligibly small for the events containing pions emitted at forward angles. Thus, the computation results depend only on the spin-triplet parts, $a^t_{\Lambda n}/a^t_{\Lambda p}$ and $r^t_{\Lambda n}/r^t_{\Lambda p}$. Angular selection of pion serves as a background reduction, and the higher momentum selection on nucleons reduces backgrounds. By taking the ratio of $\pi^0\Lambda n$ to $\pi^-\Lambda p$ spectra with several values of $a^t_{\Lambda n}/a^t_{\Lambda p}$ and $r^t_{\Lambda n}/r^t_{\Lambda p}$, we show the sensitivity of the spectra to these values.

The remainder of this article is arranged as follows. Section~\ref{sec:for} explains the method of calculating the $\Lambda N$ invariant mass spectrum. Section~\ref{sec:res} shows numerical results of the $\Lambda N$ invariant mass spectra and isospin ratios of them. Section~\ref{sec:dis} compares the results to an old experiment and examines correction of Coulomb interaction. Section~\ref{sec:sum} is devoted to a summary of this work.

%% file: formulation.tex
\section{Formulation}\label{sec:for}
\subsection{Kinematics}
The $K^- d \rightarrow \pi \Lambda N$ reaction necessitates five kinematical variables to define the momentum phase space of a three-body final state.
The $\Lambda N$ invariant mass spectrum of this reaction is calculated by integrating the square of its scattering amplitude $|\mathcal{T}|^2$ for four variables as
\begin{equation}\label{equ:CS_formula}
\frac{d\sigma}{dM_{\Lambda N}} =
\frac{M_d M_\Lambda M_N}{(2\pi)^5 4 k_\text{cm} E_\text{cm}^2} 
\int|\mathcal{T}|^2 |\bm{p}_\pi| |\bm{p}_\Lambda^*| d\Omega  d\Omega^*,
\end{equation}
where $M_d$, $M_\Lambda$, and $M_N$ are the masses of deuteron, $\Lambda$-baryon, and nucleon, respectively. Quantities defined in the center-of-mass (c.m.) frame are the initial momentum $k_\text{cm}$, the total energy $E_\text{cm}$, the momentum of the pion $\bm{p}_\pi$, and the solid angle of the pion $\Omega$. Those in the $\Lambda N$ rest frame are the momentum of the $\Lambda$-baryon $\bm{p}_\Lambda^*$, and the solid angle of $\Lambda$-baryon $\Omega^*$.

We sum up all possible spin states for each particle in the cross section. Any spin polarization is not considered for the initial deuteron, $\Lambda$-baryon nor nucleon in the final state. For deuteron having spin-1, a spin operator $S_a^\dag$ for the projection of the proton-neutron system to the spin-triplet state is written as
\begin{equation}
S_a^\dag = \frac{i}{\sqrt{2}}\sigma_a\sigma_2, \ \ (a=1,2,3),
\end{equation}
with normalizing condition $\tr[S_aS_b^\dag]=\delta_{ab}$. The square of the amplitude $|\mathcal{T}|^2$ is considered as the spin-average of the amplitude $\mathcal{T}_a$ with deuteron spin subscript $a$ as
\begin{equation}
|\mathcal{T}|^2 = \frac{1}{3} \sum_{a=1}^3 \tr[\mathcal{T}_a \mathcal{T}^\dag_a].
\end{equation}

\subsection{Scattering amplitudes}\label{subsec:amplitude}
\begin{figure}[tbp]
  \centering
  \includegraphics[width=85mm]{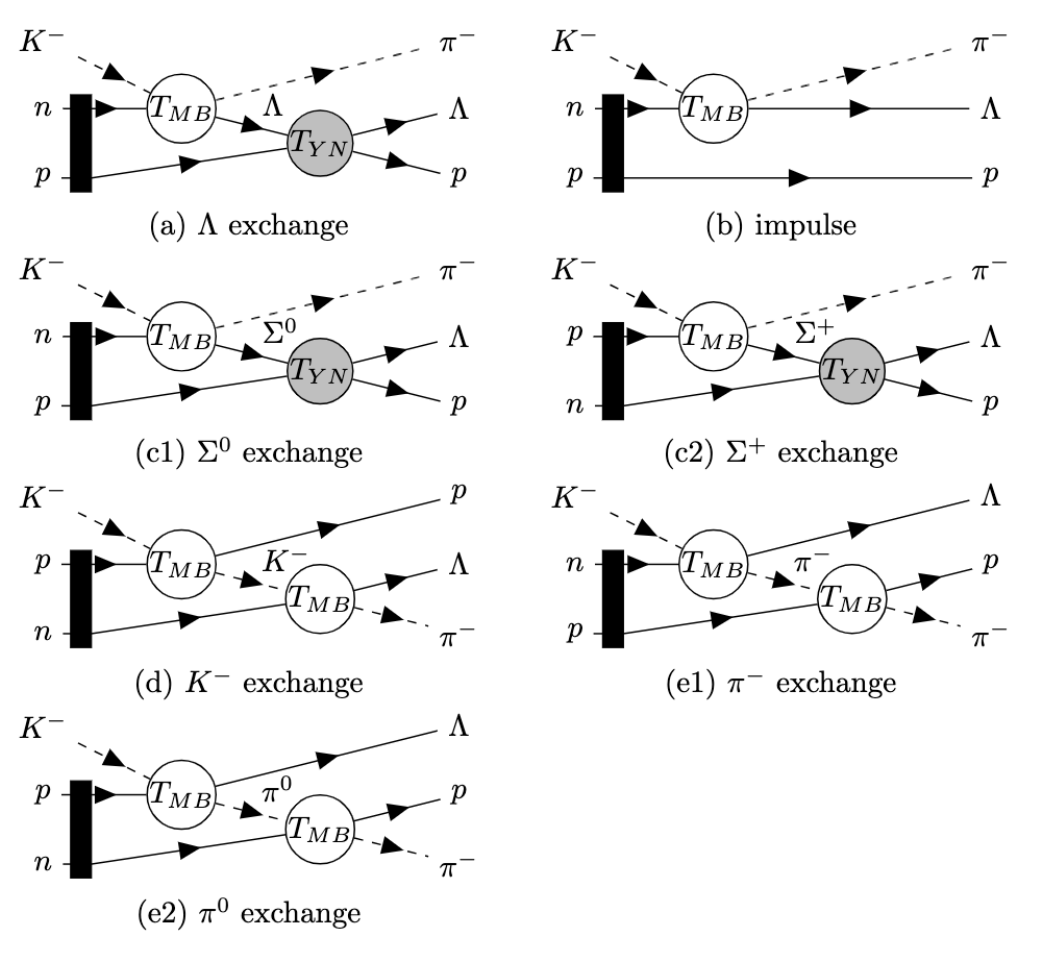}
  \caption{Feynman diagrams of the $\pi^- \Lambda p$ final state. The thick line connecting the proton and the neutron represents the deuteron. In the diagrams, $T_{MB}$ denotes the meson-baryon amplitude and $T_{YN}$ denotes the hyperon-nucleon amplitude.}
  \label{fig:p_process}
\end{figure}

\begin{figure}[htbp]
  \centering
  \includegraphics[width=85mm]{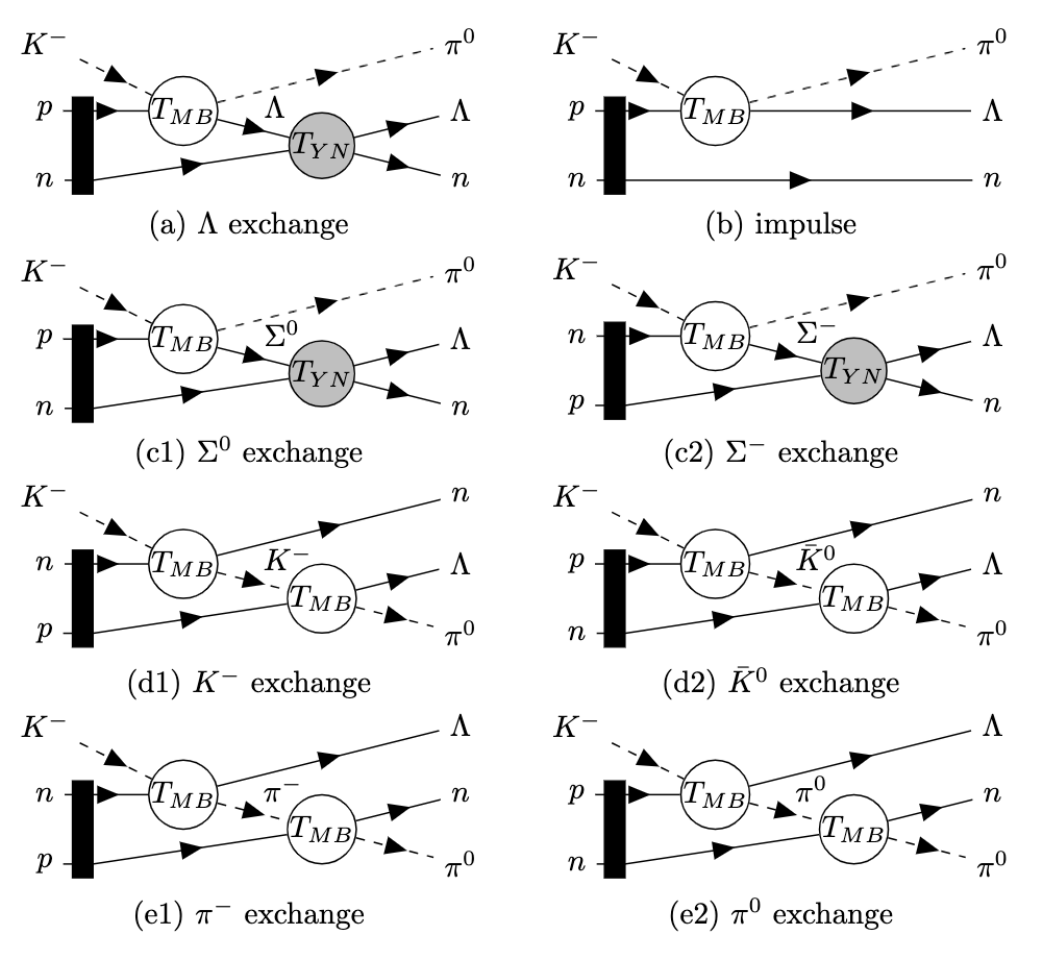}
  \caption{Feynman diagrams of the $\pi^0 \Lambda n$ final state.}
  \label{fig:n_process}
\end{figure}

In this work, the scattering amplitude $\mathcal{T}_a$ is calculated up to two-step scattering processes. Figures~\ref{fig:p_process} and \ref{fig:n_process} depict Feynman diagrams for the $K^-d\rightarrow\pi^-\Lambda p$ and $K^-d\rightarrow\pi^0\Lambda n$ reactions, respectively. Diagram (a) in Figs.~\ref{fig:p_process} and \ref{fig:n_process} represents a $\Lambda$ exchange process composed of two-step scattering of $K^-N\rightarrow\pi\Lambda$ and $\Lambda N\rightarrow\Lambda N$.  Since the process involves the final state interaction of $\Lambda N$, it is our foreground process. Two-body amplitudes $T_{MB}$ and $T_{YN}$ in diagrams denote meson-baryon and hyperon-nucleon scattering amplitudes, respectively. Diagram (b) represents an impulse process involving only one-step scattering of $K^-N\rightarrow\pi\Lambda$. Diagram (c) represents a $\Sigma$ exchange process involving two-step scattering of $K^-N\rightarrow\pi\Sigma$ and hepyron conversion $\Sigma N\rightarrow\Lambda N$. Diagram (d) represents a $\bar{K}$ exchange process involving two-step scattering of $K^-N\rightarrow \bar{K}N$ and $\bar{K}N\rightarrow\pi\Lambda$. Diagram (e) represents a $\pi$ exchange process involving two-step scattering of $K^-\Lambda\rightarrow\pi\Lambda$ and $\pi N\rightarrow\pi N$. Diagrams (b), (c), (d), and (e) are our background processes.

Next, we write down the scattering amplitudes for each process by utilizing the method of the previous study in the case of stopped kaons \cite{iizawa_2022}. Its formulation was originally described in detail in Ref.~\cite{jido_2009}. The amplitude $\mathcal{T}_a$ contains hyperon-nucleon amplitude $T_{YN}$, meson-baryon amplitude $T_{MB}$, deuteron wave function $\tilde{\varphi}(p)$ where $p$ denotes momentum fluctuation of nucleons in a deuteron, and deuteron spin matrices $S_a^\dag$. Details of these components are described in subsequent sections. Here we explain how the amplitudes are constructed by these components. We introduce the momentum of incoming kaons $\vec{p}_{K^-}$ and the momenta of outgoing particles $\vec{p}_\pi$, $\vec{p}_\Lambda$, $\vec{p}_N$. From the momentum conservation, these momenta satisfy $\vec{p}_{K^-} = \vec{p}_\pi + \vec{p}_\Lambda + \vec{p}_N$ in the deuteron rest frame. In this subsection, the physical quantities such as angles and momenta are described in the deuteron rest frame.

The amplitude of the one-step scattering process represented by diagram (b) in Figs.~\ref{fig:p_process} and \ref{fig:n_process} is described by the following formulation. If we denote the spin states of nucleon and $\Lambda$ as $\chi_N$ and $\chi_\Lambda$, respectively, the amplitude for the $K^-N\rightarrow\pi\Lambda$ process, specifying each spin state is expressed as
\begin{equation}
(\chi_\Lambda)^\dag T_{K^-N\rightarrow\pi\Lambda}\chi_N, \label{equ:kNpL}
\end{equation}
where $T_{K^-N\rightarrow\pi\Lambda}$ represents the spin matrix of the amplitude. In diagram (b), 
a meson-baryon amplitude is represented by two kinematical parameters, the invariant mass of the final $\pi\Lambda$ system $M_{\pi\Lambda}$ and the angle between the incident kaon and the emitted pion $\theta$, as $T_{K^-N\rightarrow\pi\Lambda}(M_{\pi\Lambda},\theta)$. The deuteron contribution in diagram (b) is seen in spin projection operator $S_a^\dag$ and wave function $\tilde{\varphi}$. The meson-baryon amplitude is multiplied by $S_a^\dag$ from the right because nucleon spin in the amplitude is at the right side as represented in Eq.~\eqref{equ:kNpL}, and the operator $S_a^\dag$ projects spin of one nucleon in deuteron into that of an unbound nucleon. The deuteron wave function is represented by $\tilde{\varphi}(|\vec{p}_N|)$ using the momentum of the final nucleon by momentum conservation. Therefore, the spin matrix of the amplitude $\mathcal{T}^{(I)}_a$ for the impulse process is represented by
\begin{equation}
\mathcal{T}^{(I)}_a = T_{K^-N\rightarrow\pi\Lambda}(M_{\pi\Lambda},\theta) S^\dag_a \tilde{\varphi}(|\vec{p}_N|). \label{equ:imp}
\end{equation}

For a two-step scattering process, an intermediate particle is involved, which is represented by a propagator. In the case of a hyperon exchange process depicted by diagrams (a) and (c), a hyperon propagator is utilized, which we express in a simplified relativistic form as
\begin{equation}
\frac{2M_Y}{q^2 - M_Y^2 + i\epsilon}=\frac{2M_Y}{q_0^2 -|\vec{q}|^2- M_Y^2 + i\epsilon},
\end{equation}
where $M_Y$ represents the mass of the hyperon and $q^2$ denotes the momentum transfer squared. Starting from the final state hyperon in the diagrams (a) and (c), there is the amplitude of $YN\rightarrow\Lambda N$, represented by $T_{YN}(M_{\Lambda N})$ where $M_{\Lambda N}$ is the invariant mass of the $\Lambda N$ system, an intermediate hyperon represented by propagator and integral with respect to three-vector momentum.
Energy transfer $q_0$ for hyperon exchange process is written as
\begin{equation}
q_0^{(Y)} = M_{K^-} + \left(M_N - \frac{B_d}{2}\right) - E_\pi, \label{equ:q0_hyp}
\end{equation}
where $M_N$ is the participant mass for $K^-N\rightarrow\pi\Lambda$, $E_\pi$ is energy of pion, and $B_d$ is the binding energy of deuteron. Equation~(\ref{equ:q0_hyp}) represents energy conservation at the $K^-N\rightarrow\pi Y$. In more detail, here we define $q_0$ while ignoring the correction term for the nucleon kinetic energy from the deuteron. References \cite{miyagawa_2012,jido_2013} pointed out such a correction is necessary to accurately describe the position of thresholds. Neglecting this effect would cause the threshold derived from the propagator appear at a lower energy position than its realistic location, as we would be underestimating the energy. However in our work, since we are analyzing the $\Lambda N$ invariant mass spectrum near its threshold, there is no concern that the $\Lambda N$ threshold appear at a lower position. In addition, incorporating for kinetic energy results in only little quantitative changes in the calculation. Thus, we do not consider such corrections to reduce the computational cost. The next consideration is the amplitude of $K^-N\rightarrow\pi Y$, represented by $T_{K^-N\rightarrow\pi Y}(W,\theta)$. The invariant mass of the incident kaon and one of the nucleons from deuteron $W$ is calculated by
\begin{equation}
W = \left(M_N - \frac{B_d}{2}\right) + E_{K^-}, \label{equ:W}
\end{equation}
which is from energy conservation. Equation~\eqref{equ:W} is also applicable to meson exchange processes described in subsequent sentences. The part with spin matrices, including hyperon-baryon and meson-baryon amplitudes, is divided into two terms: spin independent term $\mathcal{T}_a^1$ and spin dependent term $\mathcal{T}_a^\sigma$ as
\begin{align}
\mathcal{T}_a^1 &= T_{YN}^1(M_{\Lambda N}) T_{K^-N\rightarrow\pi Y}(W,\theta)  S^{\dag}_a, \label{equ:hyp_1}\\
\mathcal{T}_a^\sigma &= T_{YN}^\sigma(M_{\Lambda N}) \sum_b \sigma^b_{\Lambda} T_{K^-N\rightarrow\pi Y}(W,\theta) S^{\dag}_a  [\sigma^b_N]^T, \label{equ:hyp_sigma}
\end{align}
where $\sigma_\Lambda,\sigma_N$ represent spin operator for the final state baryons. The terms $T_{YN}^1$ and $T_{YN}^\sigma$ are described as linear combinations of spin singlet and triplet amplitudes,
\begin{align}
T^1_{YN} &= \frac{1}{4}(3T^t_{YN} + T^s_{YN}), \label{equ:T1}\\
T^\sigma_{YN} &=\frac{1}{4} (T^t_{YN} - T^s_{YN}). \label{equ:Tsigma}
\end{align}
Calculation details of these linear combinations are described in Sec.~\ref{sec:linear}. Utilizing the deuteron wave function, where the argument is the nucleon momentum in the deuteron determined by momentum conservation, the amplitude of the hyperon exchange process is expressed as
\begin{align}
\mathcal{T}^{(Y)}_a &= \left(\mathcal{T}_a^1 + \mathcal{T}_a^\sigma \right)\notag\\ 
&\times \int \frac{d^3\vec{q}}{(2\pi)^3} \frac{2M_Y}{q^2 - M_Y^2 + i\epsilon} \tilde{\varphi}(|\vec{q} + \vec{p}_\pi - \vec {p}_{K^-}|). \label{equ:hypex}
\end{align}

Another type of two-step scattering process, represented in diagrams (d) and (e), is the meson exchange process, which involves an intermediate meson. The amplitude of the meson exchange process has a similar structure to the hyperon exchange. The integral with respect to the three momentum of intermediate meson and the deuteron wave function are expressed in the same way as in the case of hyperon exchange. Spin matrix part of the amplitude for $\bar{K}$ exchange process expressed by diagram (d) is composed of the amplitude of $K^-N\rightarrow\bar{K}N$, the deuteron spin projection $S_a^\dag$ and the amplitude of $\bar{K}N\rightarrow\pi\Lambda$. The amplitude of $K^-N\rightarrow\bar{K}N$ is written by $T_{K^-N\rightarrow\bar{K}N}(W,\theta_1(\vec{q}))$ where $\theta_1(\vec{q})$ is the angle between the incident kaon and the intermediate kaon as $\cos\theta_1 = \hat{p}_K\cdot\hat{q}$. The amplitude of $\bar{K}N\rightarrow\pi\Lambda$ is written by $T_{\bar{K}N\rightarrow\pi\Lambda}(M_{\pi\Lambda},\theta_2(\vec{q}))$ where $\theta_2(\vec{q})$ is the angle between the intermediate kaon and the emitting pion as $\cos\theta_2 = \hat{q}\cdot\hat{p}_\pi$. Using these quantities, the amplitude of $\bar{K}$ exchange is written as
\begin{align}
\mathcal{T}^{(\bar{K})}_a &= \int \frac{d^3\vec{q}}{(2\pi)^3} \frac{1}{q^2 - M_K^2 + i\epsilon} \tilde{\varphi}(|\vec{q} + \vec{p}_N - \vec {p}_{K^-}|)\notag\\
& T_{\bar{K}N\rightarrow\pi\Lambda}(M_{\pi\Lambda},\theta_2(\vec{q})) S^{\dag}_a \left[T_{K^-N\rightarrow\bar{K}N}(W,\theta_1(\vec{q}))\right]^T. \label{equ:kex}
\end{align}
The meson-baryon amplitudes cannot be taken outside of the integral because $\theta_1$ and $\theta_2$ depend on $\vec{q}$. From the energy conservation in the same manner as Eq.~\eqref{equ:q0_hyp}, $q_0$ is determined as
\begin{equation}
q_0^{(\bar{K})} = M_{K^-} + \left(M_N - \frac{B_d}{2}\right) - E_N.
\end{equation}
The amplitude for $\pi$ exchange process is written in the same way of $\bar{K}$ exchange process as
\begin{align}
\mathcal{T}^{(\pi)}_a &= \int \frac{d^3q}{(2\pi)^3}\frac{1}{q^2 - M_\pi^2 + i\epsilon} \tilde{\varphi}(|\vec{q} + \vec{p}_\Lambda   - \vec {p}_{K^-}|)\notag\\
& T_{K^-N\rightarrow\pi\Lambda}(W,\theta_1(\vec{q})) S^{\dag}_a \left[T_{\pi N\rightarrow\pi N}(M_{\pi N},\theta_2(\vec{q}))\right]^T \label{equ:piex}
\end{align}
where $\theta_1,\theta_2$ are the same formulation of kaon exchange process and $q_0$ is calculated as
\begin{equation}
q_0^{(\pi)} = M_{K^-} + \left(M_N - \frac{B_d}{2}\right) - E_\Lambda.
\end{equation}

In Figs.~\ref{fig:p_process} and \ref{fig:n_process}, in order to separate the amplitude for each charged state, such as $\pi^-$ exchange and $\pi^0$ exchange, it is crucial to account for the relative signs of these amplitudes within each pair of nucleons in the deuteron. Since the deuteron is an isospin singlet particle, the distinct charged states represented by the permutation of proton and neutron must be antisymmetrized. For instance, when considering $\pi$ exchange, the amplitude is calculated by
\begin{equation}
\mathcal{T}^{(\pi)}_a = \frac{1}{\sqrt{2}}\left[\mathcal{T}^{(\pi^-)}_a - \mathcal{T}^{(\pi^0)}_a\right].
\end{equation}

\subsection{Hyperon-Nucleon amplitudes}
Since hyperon and nucleon are both spin-1/2 particles, there are two spin states in the hyperon-nucleon system, spin-singlet and apin-triplet. Each amplitude is parametrized by effective range expansion:
\begin{align}
T^s_{YN} = \mathcal{N} \frac{1}{-\frac{1}{a^s_{YN}} + \frac{1}{2} r^s_{YN} p^{*2}_Y - i p^*_Y},\\
T^t_{YN} = \mathcal{N} \frac{1}{-\frac{1}{a^t_{YN}} + \frac{1}{2} r^t_{YN} p^{*2}_\Lambda - i p^*_Y} \label{equ:eff_range_t},
\end{align}
where $a_{YN}$, $r_{YN}$ and $p^*_Y$ are the scattering length, the effective range and the momentum of the $YN$ system in the $YN$ rest frame, respectively. Kinematic factor $\mathcal{N}$ is expressed by
\begin{equation}
\mathcal{N} = - \frac{8\pi M_{YN}}{\sqrt{(2M_{Y_i})(2M_{N_i})(2M_{Y_f})(2M_{N_f})}}.
\end{equation}
The amplitude of $\Lambda p\rightarrow\Lambda p$ is calculated by using the central values of experimental data for the $\Lambda p$ system; $a^s_{\Lambda p} = -2.43^{+0.16}_{-0.25}$ fm, $a^t_{\Lambda p} = -1.56^{+0.19}_{-0.22}$ fm, $r^s_{\Lambda p} = 2.21^{+0.16}_{-0.36}$ fm, $r^t_{\Lambda p} = 3.7^{+0.6}_{-0.6}$ fm~\cite{budzanowski_2010}. For the spin-triplet state of $\Lambda n$ system, we define the ratios $a^t_{n/p}$ and $r^t_{n/p}$ as
\begin{align}
a^t_{n/p} &= a^t_{\Lambda n}/a^t_{\Lambda p}, \label{equ:ratioat}\\
r^t_{n/p} &= r^t_{\Lambda n}/r^t_{\Lambda p}. \label{equ:ratiort}
\end{align}
Computations of the invariant mass spectra for the $K^-d\rightarrow\pi^0\Lambda n$ are performed with several values of these ratios. The spin-singlet component contributes very little due to the kinematic selection for pions going forward, and this is seen in Eq.~\eqref{equ:spin_sel} of Appendix~\ref{sec:spin_mat}. Therefore, here we define only the ratio for the spin-triplet, and do not consider the spin-singlet component. 
In Sec.~\ref{subsec:ratio}, we take the ratio of the spectra for the reactions $K^-d\rightarrow\pi^-\Lambda p$ and $K^-d\rightarrow\pi^0\Lambda n$, and investigate the sensitivity of this ratio to those defined in Eqs.~\eqref{equ:ratioat} and \eqref{equ:ratiort}.

The $\Sigma N\rightarrow\Lambda N$ transition amplitude $T_{\Sigma N \rightarrow \Lambda N}$ is calculated by unitarity of coupled channels of $\Sigma N$ and $\Lambda N$. Coupled channel amplitudes are formulated in Ref.~\cite{cohen_2004} and its application to $\Lambda N\rightarrow\Sigma N$ is written in Ref.~\cite{iizawa_2022}. We use two theoretical calculations, $a_{\Sigma N}=1.68-2.35i$ fm \cite{rijken_1999} called NSC97f, ${\Sigma N}=-3.83-3.01i$ fm \cite{haidenbauer_2005} called J\"ulich'04. The calculation results are insensitive of the choice of the parameters around the $\Lambda N$ threshold because the effect of the $\Sigma$ exchange process is negligibly small.

\subsection{Meson-Baryon amplitudes}
The partial wave decomposition of the meson-baryon amplitude is expressed by the following formula:
\begin{align}
T_{ij} = &\sum_{l=0}^\infty \left[(l+1)T_{ij}^{l+} + lT_{ij}^{l-}\right]P_l(\hat{k}_j\cdot\hat{k}_i) \notag\\
& - i\vec{\sigma}\cdot(\hat{k}_j\times\hat{k}_i)\sum_{l=0}^\infty (T_{ij}^{l+} - T_{ij}^{l-})P'_l(\hat{k}_j\cdot\hat{k}_i)\label{equ:partial},
\end{align}
where $l$ denotes orbital angular momentum, and $P_l$ is Legendre polynomials. With considering the baryon spin 1/2, $V^{l+}$ and $V^{l-}$ denote amplitudes of total angular momentum $l+1/2$ and $l-1/2$, respectively.

For pion-nucleon ($\pi N$) scattering, we use the amplitudes obtained by the data of partial wave analysis \cite{workman_2012}. In the data, there are $\pi N$ partial wave amplitudes, $T^{0}_{\pi N}(W), T^{1+}_{\pi N}(W), T^{1-}_{\pi N}(W)$. Amplitudes for each charged state of $\pi N$ is obtained by multiplying Clebsch-Gordan coefficients.

We use the chiral unitary amplitude with partial wave expansion up to p-wave for the $\bar{K}N$ amplitude $T_{\bar{K}N}$ developed in the article \cite{jido_2002}. The interaction kernel is derived from leading-order Lagrangian of chiral perturbation theory. The leading-order Lagrangian consists of two terms, Weinberg-Tomozawa term and Born term. Weinberg-Tomozawa term for $i$ to $j$ channels is written as
\begin{equation}
V^\text{(WT)}_{ij} = -\frac{C_{ij}}{4f^2}\bar{u}(p_j)(\cancel{k}_i + \cancel{k}_j)u(p_i),
\end{equation}
where $f$ is a typical decay constant of octet mesons, $C_{ij}$ is a channel coefficient of chiral perturbation theory, its values are listed in Ref.~\cite{oset_1998}. $k_i$ and $p_i$ denote the momenta of meson and baryon for the meson-baryon state $i$, respectively. The other leading-order term, Born term, is written as
\begin{equation}
V^\text{(Born)}_{ij} = D_{iY} D_{Yj}\bar{u}(p_j) \cancel{k}_j\gamma_5\frac{1}{(\cancel{p}_i+\cancel{k}_i) - \tilde{M}_Y} \cancel{k}_i\gamma_5 u(p_i),
\end{equation}
where $Y$ denotes channel index of the propagating hyperon, $\Lambda$ and $\Sigma$, and their bare masses are written by $\tilde{M}$. The constants $D_{ik}$ are channel coefficients described in Ref.~\cite{jido_2002}. We also consider a hyperon resonance $\Sigma^*$, with spin 3/2. This state does not belong to the octet baryon. We use spin-flavor SU(6) symmetry to determine the coupling constants and spin transition operator $\vec{\mathcal{S}}
$ to express the spin transition in $MB\rightarrow\Sigma^*$ and which satisfies the following property:
\begin{equation}
\begin{aligned}
\mathcal{S}_i^\dag P_{3/2} \mathcal{S}_j &= \sum_{M=-3/2}^{3/2}\mathcal{S}_i^\dag \ketbra{1,\frac{1}{2},\frac{3}{2},M}\mathcal{S}_j\\
&= \frac{2}{3}\delta_{ij}-\frac{i}{3}\epsilon_{ijk}\sigma_k,
\end{aligned}
\end{equation}
where $P_{3/2}$ is the projection operator of spin-3/2 system. Using Dirac spinor for baryon
\begin{equation}
u(p) = \sqrt{\frac{p_0 + M}{2p_0}}
\begin{pmatrix}
\chi \\ \frac{\vec{\sigma}\cdot\vec{p}}{p_0 + M}\chi
\end{pmatrix},
\end{equation}
with 2-spinor $\chi$, the spin matrix part of the sum of two kernels, $V_{ij}$ is expanded as follows for each order of the meson's momentum $\vec{k}$ in the center mass frame:
\begin{align}
V_{ij} = &-\frac{C_{ij}}{4f^2} \alpha_i \alpha_j(2W - M_i - M_j) \notag\\
&- \frac{C_{ij}}{4f^2} \alpha_i \alpha_j \left(\frac{1}{\beta_i} + \frac{1}{\beta_j}\right)\left(\vec{k}_i\cdot\vec{\sigma}\right)\left(\vec{k}_j\cdot\vec{\sigma}\right) \notag\\
&+ \sum_{Y=\Lambda,\Sigma} \biggl[\frac{D_{iY}D_{Yj}}{W-\tilde{M}_Y}\left(\vec{k}_i\cdot\vec{\sigma}\right)\left(\vec{k}_j\cdot\vec{\sigma}\right) \notag\\
&\quad\times\biggl(1+\frac{k_i^0}{M_i}\biggr)\biggl(1+\frac{k_j^0}{M_j}\biggr)\biggr] \notag\\
&+ \frac{D_{i\Sigma^\star}D_{\Sigma^\star j}}{W - \tilde{M}_{\Sigma^\star}} (\vec{\mathcal{S}}\cdot\vec{k}_i) (\vec{\mathcal{S}}^\dag\cdot\vec{k}_j) + \mathcal{O}(k^2),
\end{align}
where
\begin{align}
\alpha_i &= \sqrt{\frac{E_i + M_i}{2M_i}}, \\
\beta_i &= E_i + M_i.
\end{align}
This kernel $V_{ij}$ can be decomposed into partial wave amplitude in Eq.~(\ref{equ:partial}).
If we consider up to p-wave ($l=0,1$), the s-wave kernel $V^{0}_{ij}$ and p-wave kernel components $V^{1+}_{ij},V^{1-}_{ij}$ are written as
\begin{align}
V^{0}_{ij} &= -\frac{C_{ij}}{4f^2} \alpha_i \alpha_j(2W - M_i - M_j), \\
V^{1+}_{ij} &= k_i k_j\frac{1}{3} \frac{D_{i\Sigma^\star}D_{\Sigma^\star j}}{W-\tilde{M}_{\Sigma^\star}},\\
V^{1-}_{ij} &= k_i k_j\biggl[-\frac{C_{ij}}{4f^2} \alpha_i \alpha_j \left(\frac{1}{\beta_i} + \frac{1}{\beta_j}\right) \notag\\
&\qquad + \frac{D_{i\Lambda}D_{\Lambda j}}{W - \tilde{M}_{\Lambda}}\biggl(1+\frac{k_i^0}{M_i}\biggr)\biggl(1+\frac{k_j^0}{M_j}\biggr) \notag\\
&\qquad + \frac{D_{i\Sigma}D_{\Sigma j}}{W - \tilde{M}_{\Sigma}}\biggl(1+\frac{k_i^0}{M_i}\biggr)\biggl(1+\frac{k_j^0}{M_j}\biggr)\biggr],
\end{align}
respectively.

Unitarization of these amplitude kernels are done by following equations with a loop function $\mathcal{G}_i$,
\begin{align}
T^{0}_{ij} &= V^0_{ij} + V^0_{ik}\mathcal{G}_kT_{kj} \label{equ:uni_s},\\
T^{1+}_{ij} &= V^{1+}_{ij} + V^{1+}_{ik}\mathcal{G}_kT_{kj} \label{equ:uni_p},\\
T^{1-}_{ij} &= V^{1-}_{ij} + V^{1-}_{ik}\mathcal{G}_kT_{kj} \label{equ:uni_m}.
\end{align}
The loop function $\mathcal{G}_i$ is obtained by a loop integral for invariant mass $W$ with subtraction constants $a_{i}(\mu)$ for the regularization scale $\mu$,
\begin{align}
\mathcal{G}_i(W) =& 2iM_i \int\frac{d^4q}{(2\pi)^4}\frac{1}{(p-q)^2-M_i^2+i\epsilon}\frac{1}{q^2-m_i^2+i\epsilon}\notag\\
=& \frac{M_i}{8\pi^2}\Bigg[a_i(\mu) + \ln(\frac{M_i^2}{\mu^2}) \notag\\
&+ \frac{m_i^2-M_i^2+W^2}{2W^2}\ln(\frac{m_i^2}{M_i^2}) \notag\\
&+ \frac{p_\text{cm}}{W}\Big\{\ln(W^2-(M_i^2-m_i^2))+2p_\text{cm}W \notag\\
&+ \ln(W^2 + (M_i^2 - m_i^2) + 2p_\text{cm}W) \notag\\
&- \ln(-W^2 + (M_i^2 - m_i^2) + 2p_\text{cm}W) \notag\\
&- \ln(-W^2 - (M_i^2 - m_i^2) + 2p_\text{cm}W) \Big\}\Bigg],
\end{align}
where $m_i$ and $M_i$ are the masses of meson and baryon, respectively. $p_\text{cm}$ is the momentum in the center mass frame for meson-baryon system. Lastly we write down the values of the model parameters, which were determined in Ref.~\cite{jido_2002}, the decay constant $f=104.4$ MeV, the regularization scale $\mu=630$ MeV, bare masses of hyperons in the Born terms $\tilde{M}_\Lambda=1069$ MeV, $\tilde{M}_{\Sigma^0}=1195$ MeV, $\tilde{M}_{\Sigma^-}=1201$ MeV, $\tilde{M}_{\Sigma^{*0}}=1413$ MeV, $\tilde{M}_{\Sigma^{*-}}=1420$ MeV and subtraction constants $a_i(\mu)$ as
\begin{align}
a_{\bar{K}N} = -1.75, \ \ &a_{\pi\Sigma} = -2.10, \ \ a_{\pi\Lambda} = -1.62\notag\\
a_{\eta\Lambda} = -2.56, \ \ &a_{\eta\Sigma} = -1.54, \ \ a_{K\Xi} = -2.67.
\end{align}
With these values of the bare masses, the physical masses are reproduced.

\subsection{Deuteron wave function}
We use deuteron wave function $\tilde{\varphi}$ obtained by the CD-Bonn potential \cite{machleidt_2001} in momentum space, which is given in the deuteron rest frame (the laboratory frame). The s-wave part of the wane function is written as
\begin{equation}
  \tilde{\varphi}(p) = N\sum_{i=1}^{11} \frac{C_i}{p^2 + m_i^2}, \label{equ:deuteron}
\end{equation}
where $N$ is a normalization factor, and the values of $C_i$ and $m_i$ are listed in Ref.~\cite{machleidt_2001}. The amplitudes of all processes have the deuteron wave function to include momentum fluctuation of nucleons in deuteron. $\tilde{\varphi}(p)$ has a maximum at $p=0$, and this implies that the proton and neutron most probably have zero momenta and the probability of these particles having higher momentum fluctuation inside the deuteron is strongly suppressed.

In order to use the deuteron wave function given in the deuteron rest frame, all quantities in the c.m. frame are transformed into the deuteron rest frame when integration with respect to the kinematical valuables are performed. Whole amplitudes $\mathcal{T}_a$ are given as a Lorentz-invariant form, however their components do not necessarily posses invariance. We employ Lorentz transformation for the meson-beryon amplitudes with partial-wave expansion referencing \cite{yamagata-sekihara_2013}. When transforming from the c.m. frame to the laboratory frame, the structure of amplitudes remains unchanged, but the scale varies slightly. The momenta of particles in amplitudes are written in the deuteron rest frame.

%% file: results.tex
\section{Results}\label{sec:res}
This section shows the numerical results of the $\Lambda N$ invariant mass spectra. Initially, the outcomes for the $K^-d\rightarrow\pi^-\Lambda p$ reaction are displayed without background reductions. Subsequently, background reduction is introduced by selecting the phase space of the final state. Towards the conclusion of this section, the ratio between the spectra for the two reactions, $K^-d\rightarrow\pi^-\Lambda p$ and $K^-d\rightarrow\pi^0\Lambda n$, is shown. We perform calculations with several values of $a^t_{n/p}$ and $r^t_{n/p}$ to examine the sensitivity of the spectra to these parameters. The incident momenta of kaons are set at 1000 MeV/c throughout this section.

\subsection{$\Lambda N$ invariant mass spectrum}
The value of the $\Lambda N$ invariant mass spectrum are calculated by integration of the transition probability in the phase space of the final state denoted by Eq.~\eqref{equ:CS_formula}, which is parametrized by the solid angle of pion in the center-of-mass frame $d\Omega=d\cos\theta d\varphi$ and the polar angle of $\Lambda$ in the $\Lambda N$ rest frame $\theta^*$, as
\begin{align}
\frac{d\sigma}{dM_{\Lambda N}} =& \frac{M_d M_\Lambda M_N}{(2\pi)^4 4 k_{cm} E_{cm}^2} |\bm{p}_\pi| |\bm{p}_\Lambda^*| \notag\\
&\times \int_{-1}^1 d\cos\theta \int_0^{2\pi} d\varphi \int_{-1}^1 d\cos\theta^* |\mathcal{T}|^2.\label{equ:cross_section}
\end{align}
The azimuthal angle symmetry of $\Lambda$ in the $\Lambda N$ rest frame is utilized in the integration process.

\begin{figure}[tb]
\centering
\includegraphics[width=0.45\textwidth]{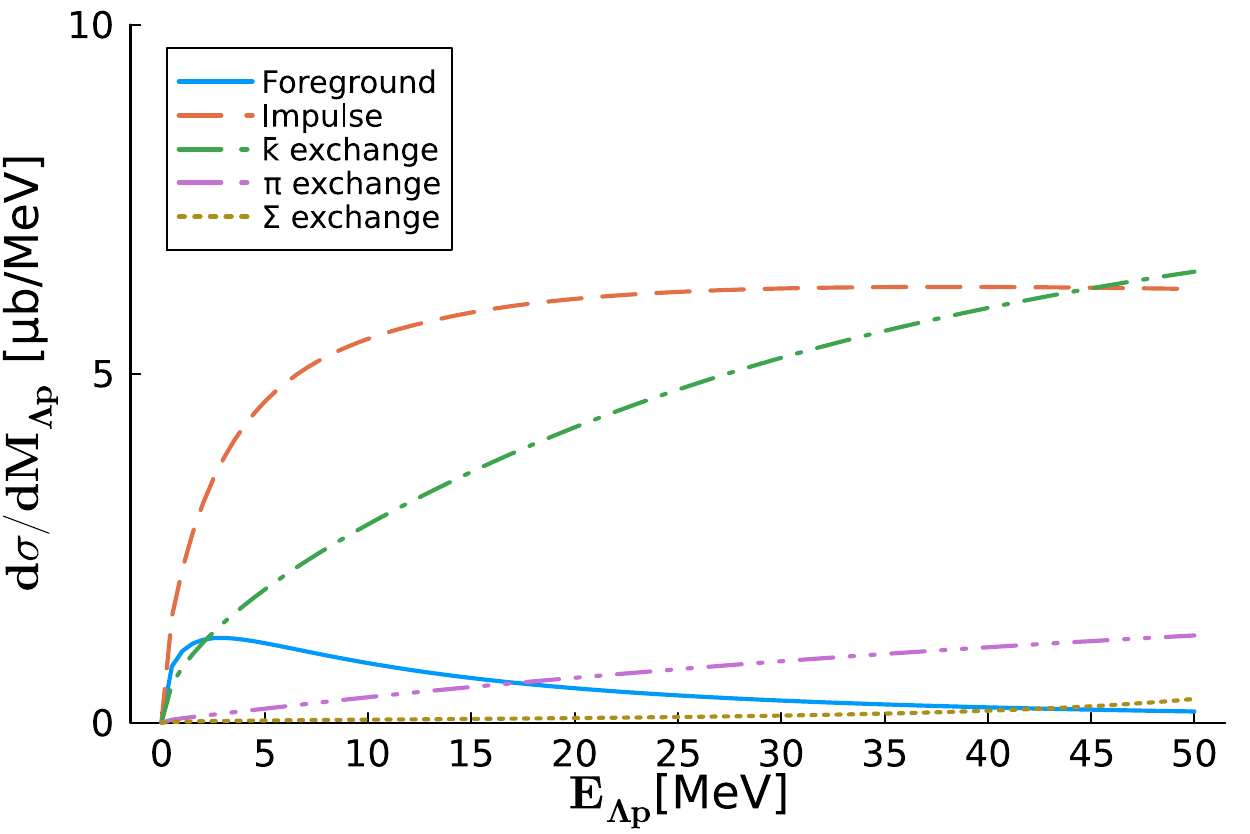}
\caption{The $\Lambda N$ invariant mass spectra for each process of the $K^- d \rightarrow$ $\pi^-\Lambda p$ reaction with incident momenta of kaons at 1000 MeV/c. Solid, dashed, dash-dotted, dash-dash-dotted, and dotted lines show foreground, impulse, $\bar{K}$ exchange, $\pi$ exchange, and $\Sigma$ exchange processes, respectively.}
\label{fig:CS_p}
\end{figure}

Figure \ref{fig:CS_p} shows the $\Lambda N$ invariant mass spectra of each diagram for the $K^-d \rightarrow \pi^-\Lambda p$ reaction. Each line represents the spectrum calculated for a specific process. For instance, the square amplitude of the impulse process is calculated by $|\mathcal{T}^{(I)}|^2 = \frac{1}{3} \tr[\mathcal{T}^{(I)}_a\mathcal{T}^{(I)\dag}_a]$. The horizontal axis represents the excitation energy $E_{\Lambda p} = M_{\Lambda p} - (M_\Lambda + M_p)$ measured from the mass threshold. We find that the foreground process exhibits a peak around 2 MeV. The impulse process dominates over other terms across most regions, and the $\bar{K}$ exchange becomes significant enough to surpass the foreground at around 3 MeV. The $\pi$ exchange gradually increases and surpasses the foreground at approximately 17 MeV. On the other hand, the contribution from $\Sigma$ exchange remains negligibly small, almost aligning with the horizontal axis. Due to the large contamination from backgrounds, it seems to be hard to extract the $\Lambda N$ interaction from the spectrum.

\subsection{Angular selections of pions}
\begin{figure}[tb]
\centering
\includegraphics[width=0.48\textwidth]{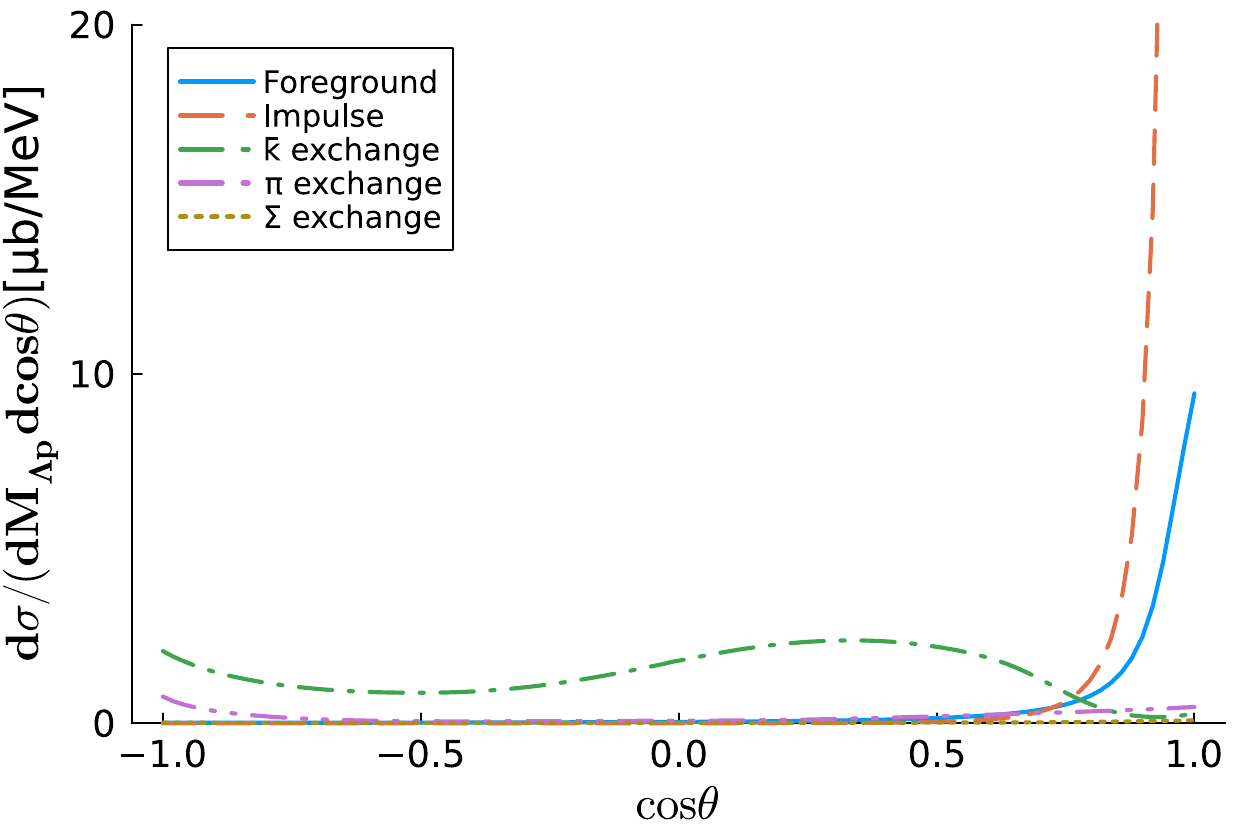}
\caption{Angular dependence of the differential cross section for each process of the $K^-d\rightarrow\pi^-\Lambda p$ reaction. The excitation energy $E_{\Lambda p}$ is fixed at 10 MeV. The angle of pion $\theta$ is calculated in the c.m. frame. Each line is plotting the contribution of each process as Fig.~\ref{fig:CS_p}.}
\label{fig:ang_dep}
\end{figure}

From Eq.~(\ref{equ:cross_section}), the spectrum is obtained by integrating the scattering amplitude across the phase space. The suppression of background processes, such as impulse, $\bar{K}$ exchange, and $\pi$ exchange processes, is achieved by constraining the phase space. We first focus on the angle of the final state pions, denoted by $\theta$ in Eq.~(\ref{equ:cross_section}). This angle represents the direction of the pion concerning the incident kaon direction. To identify a favorable range of $\theta$ for background reduction, the angular dependence of each process are plotted. Figure \ref{fig:ang_dep} depicts the $\theta$ dependence of each process at an excitation energy of $E_{\Lambda p} = 10$ MeV. The spectra are obtained by integrating $\varphi$ and $\theta^*$. Both the foreground and impulse processes exhibit maximum values at $\cos\theta=1$, which is the forward direction of the pion. Conversely, the contributions of $\bar{K}$ exchange and $\pi$ exchange are distributed over a relatively wider range of angles compared to the foreground and impulse processes.

The angular distributions of foreground and impulse processes are interpretable from the behaviors of the scattering amplitudes (\ref{equ:imp}) and (\ref{equ:hypex}) through the deuteron wave function $\tilde{\varphi}$. Since $\tilde{\varphi}(p)$ has a maximum at $p=0$, the differential cross section has larger value in the region of the phase space where the argument of the deuteron wave function has smaller value. Equation (\ref{equ:imp}) shows that it is the momentum of the final state neutron $|\vec{p}_N|$. The region where $|\vec{p}_N|$ has a lower value corresponds to that of pion going to the forward direction, $\cos\theta\sim1$. Because of the momentum conservation, the particles in the final state share the momentum of the incident kaon. When the pion carry higher momentum to the forward direction, the momentum of nucleon $|\vec{p}_N|$ is smaller. Similar argument is applicable to the foreground process close to the threshold. The deuteron wave function in Eq.~(\ref{equ:hypex}) is written by $\tilde{\varphi}(|\vec{q}+\vec{p}_\pi-\vec{p}_{K^-}|)$. When the pion goes to the forward direction, resulting in a small value for $|\vec{p}_\pi-\vec{p}_{K^-}|$. 

\begin{figure}[tb]
\centering
\includegraphics[width=0.48\textwidth]{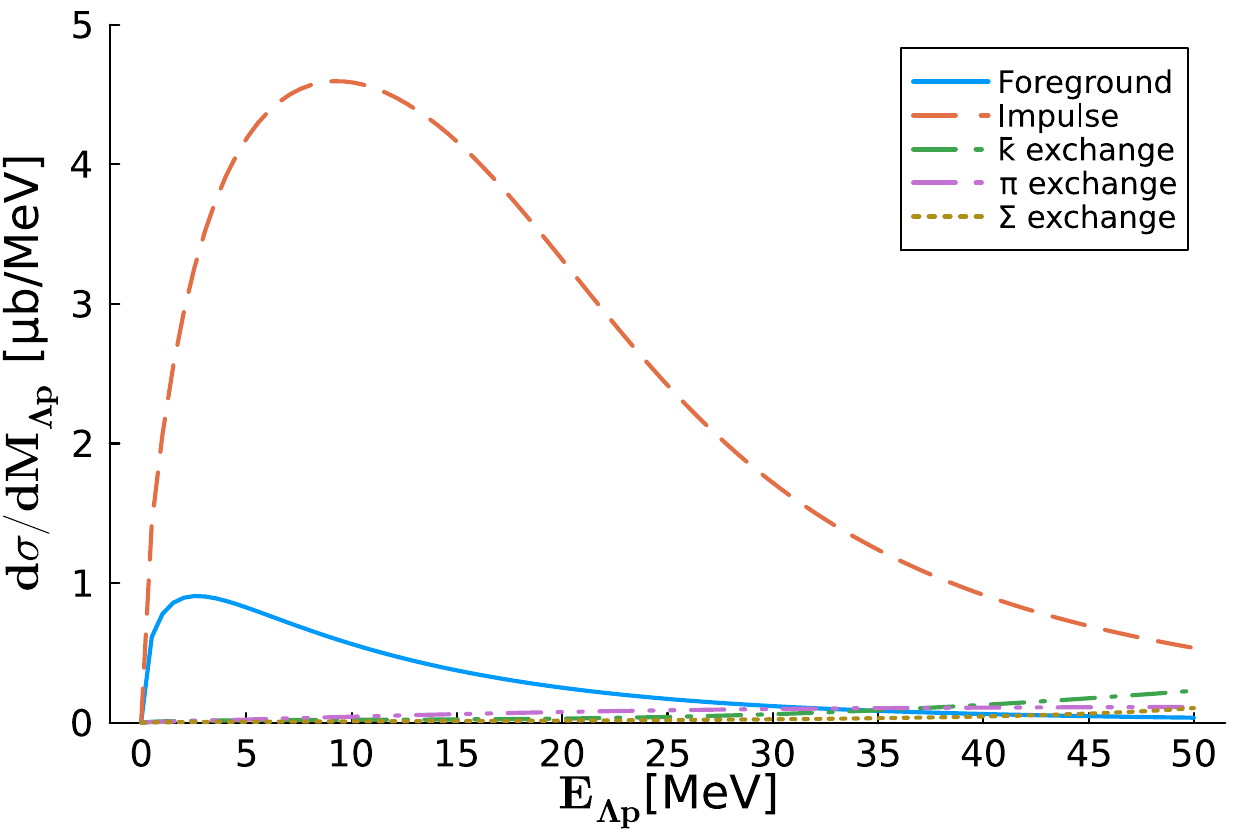}
\caption{The $\Lambda p$ invariant mass spectra for each process of the $K^-d\rightarrow\pi^-\Lambda p$ reaction with an angular selection of the pion, $0.9\le\cos\theta\le1.0$. Each line is plotting contribution of each process as Fig.~\ref{fig:CS_p}.}
\label{fig:ang_lim}
\end{figure}

Figure \ref{fig:ang_lim} displays the $\Lambda p$ invariant mass spectra for the $K^-d\rightarrow\pi^-\Lambda p$ reaction with integrating forward angles of the pion, $0.9\le\cos\theta\le1.0$. This figure illustrates a reduction of background processes involving meson changes, such as $\bar{K}$ exchange and $\pi$ exchange. Foreground and impulse processes exhibit larger values compared to $\bar{K}$ exchange and $\pi$ exchange processes. Although the contribution of the remaining background, the impulse process, gets smaller above the energy $E_{\Lambda p}=10$ MeV than the result without the angular selection depicted in Fig.~\ref{fig:CS_p}, the suppression of the impulse process with the angular selection is inadequate near the $\Lambda p$ mass threshold.

Regarding the angular selection, the explanation of spin selection within the $\Lambda N$ interaction is outlined. Both spin-singlet and spin-triplet are possible for the $\Lambda N$ interaction. However, as observed from Eq.~(\ref{equ:spin_sel}) in Sec.~\ref{subsec:hyperon}, when focusing on the kinematics where the direction of the pion momentum is forward, the contribution of spin-singlet component diminishes considerably. This predominance of the spin-triplet component is termed spin selection. In the computation of the $\Lambda n$ invariant mass spectra for the $K^-d\rightarrow\pi^0\Lambda n$ reaction, only the sensitivity of the spin triplet is discussed with considering several values of $a^t_{n/p}$ and $r^t_{n/p}$.

\subsection{Momentum selection of nucleon}
The constraint on the nucleon momentum is another selection of the phase space, and effectively reduces the contribution of the impulse process. As demonstrated in the preceding section, Eq.~\eqref{equ:imp} indicates that a region with higher nucleon momentum, evident in the deuteron wave-function $\varphi(|\vec{p}_N|)$, is more favorable for minimizing the impulse process. The region where the nucleon momentum exceeds a specified value $|\vec{p_{N}}| > p_\text{cut}$ is considered. From the width of the deuteron wave function, we choose a value as $p_\text{cut}=150$ MeV/c.

\begin{figure}[tb]
\centering
\includegraphics[width=0.48\textwidth]{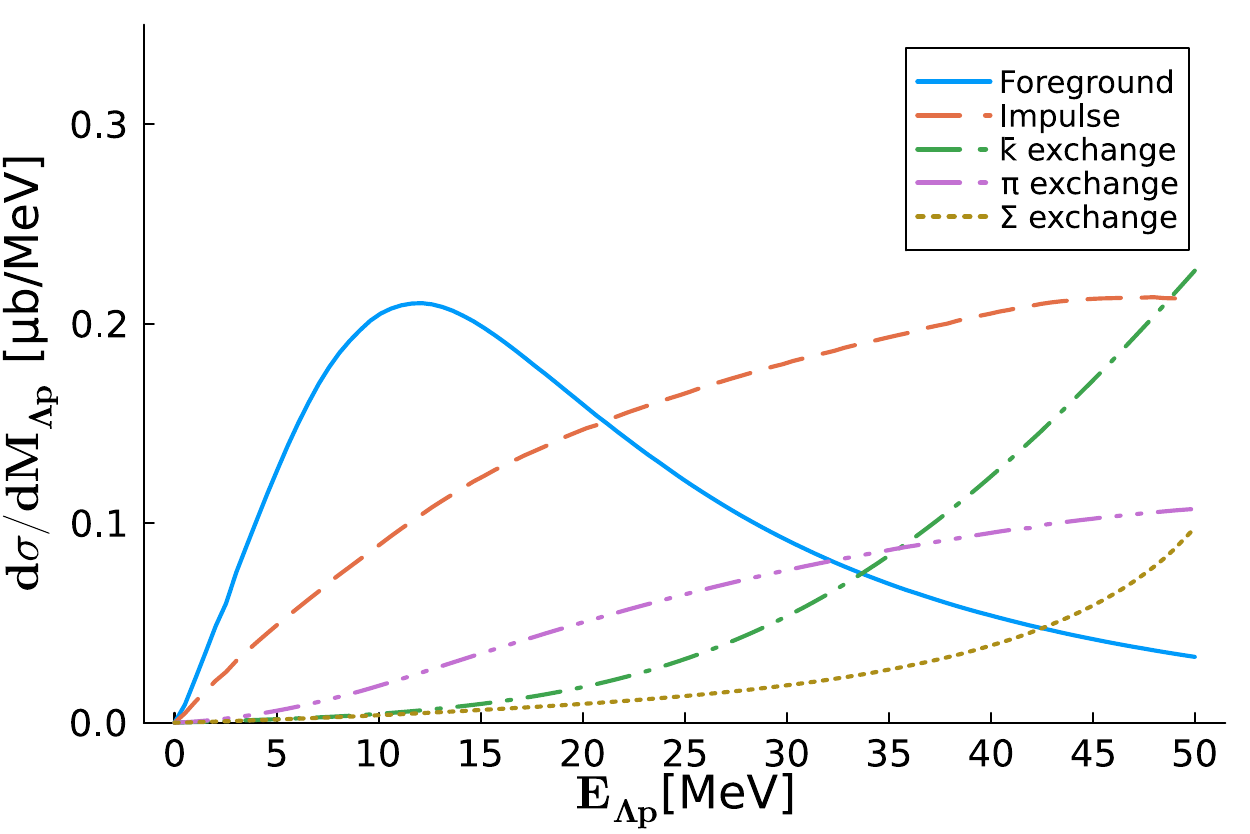}
\caption{The $\Lambda p$ invariant mass spectra for each process of the $K^-d\rightarrow\pi^-\Lambda p$ reaction with two selections, $0.9\le\cos\theta\le 1.0$  and $p_\text{cut}=150$ MeV/c. Each line is plotting contribution of each process as Fig.~\ref{fig:CS_p}.}
\label{fig:p_moment}
\end{figure}

Figure \ref{fig:p_moment} shows the numerical result with selection $0.9\le\cos\theta\le1.0$ and $p_\text{cut}=150$MeV/c. As one sees in Fig.~\ref{fig:p_moment}, the foreground process has the largest contribution up to around $E_{\Lambda p} = 20$ MeV which indicates effective suppression of the background processes. A comparison between Figs.~\ref{fig:ang_lim} and \ref{fig:p_moment} reveals notable alterations resulting from the nucleon momentum constraint on the impulse and foreground processes. There is a significant decrease in the contribution of the impulse and about three quarter reduction in the foreground compared to Fig.~\ref{fig:ang_lim}. The peak of the foreground shifts to approximately 12 MeV from its original position around 2 MeV. Furthermore, the overall magnitude of the spectrum decreases approximately to one-tenth of its previous value.

\subsection{Cross section ratio of $\pi^-\Lambda p$ and $\pi^0\Lambda n$}\label{subsec:ratio}
In previous sections, the amplitudes have been computed for each process separately. From this point forward, we calculate the sum of amplitudes for each process, as expressed in the following equation,
\begin{equation}
\mathcal{T}_a = \mathcal{T}^{(\Lambda)}_a + \mathcal{T}^{(I)}_a + \mathcal{T}^{(\bar{K})}_a + \mathcal{T}^{(\pi)}_a + \mathcal{T}^{(\Sigma)}_a.
\end{equation}

Here we compute the ratio of $\Lambda N$ invariant mass spectra between $\pi^-\Lambda p$ and $\pi^0\Lambda n$ to examine the sensitivity to $a_{n/p}^t$ and $r_{n/p}^t$. The method of taking the ratio is based on isospin symmetry of the final states, $\ket{\pi\Lambda N}_{I_3=-1/2}$. With the isopin decomposition, two isospin states of $\ket{\pi\Lambda N}_{I_3=-1/2}$ in the particle basis, $\ket{\pi^-\Lambda p}$ and $\ket{\pi^0\Lambda n}$, are represented using Clebsch-Gordan coefficients as
\begin{align}
\ket{\pi^-\Lambda p} &= -\sqrt{\frac{2}{3}}\ket{\pi\Lambda N,\frac{1}{2}} + \frac{1}{\sqrt{3}}\ket{\pi\Lambda N,\frac{3}{2}}\label{equ:pi-p},\\
\ket{\pi^0\Lambda n} &= \frac{1}{\sqrt{3}}\ket{\pi\Lambda N,\frac{1}{2}} + \sqrt{\frac{2}{3}}\ket{\pi\Lambda N,\frac{3}{2}}\label{equ:pi0n},
\end{align}
where $\ket{\pi\Lambda N,\frac{1}{2}}$ and $\ket{\pi\Lambda N,\frac{3}{2}}$ denote the $\pi\Lambda N$ states with total isospin $I=1/2$ and $I=3/2$, respectively. Since the initial state of $\ket{K^-d}$ has isospin $I=1/2$, isospin symmetry claims that the ratio of the amplitudes between each final state $\mathcal{T}_{\pi^0\Lambda n}$ and $\mathcal{T}_{\pi^-\Lambda p}$ should be fixed as
\begin{equation}
\mathcal{T}_{\pi^0\Lambda n} / \mathcal{T}_{\pi^-\Lambda p} = -\frac{1}{\sqrt{2}}.
\end{equation}
The cross section ratio of two processes are defined by
\begin{equation}
R = 2 \frac{\sigma_{\pi^0\Lambda n}}{\sigma_{\pi^-\Lambda p}},
\end{equation}
where $\sigma$ denotes the cross section. The ratio would satisfy $R = 1$ if isospin symmetry were exact. The actual situation is that isospin symmetry is an approximate symmetry, thus $R$ deviates from 1. When computing the ratio of the $\Lambda N$ invariant mass spectra across various values of $a_{n/p}$ and $r_{n/p}$, the deviations illustrate the sensitivity of $R$ to alterations in the values of scattering lengths and effective ranges.

\begin{figure*}[bt]
\begin{center}
\begin{subfigure}{0.48\textwidth}
\includegraphics[width=\textwidth]{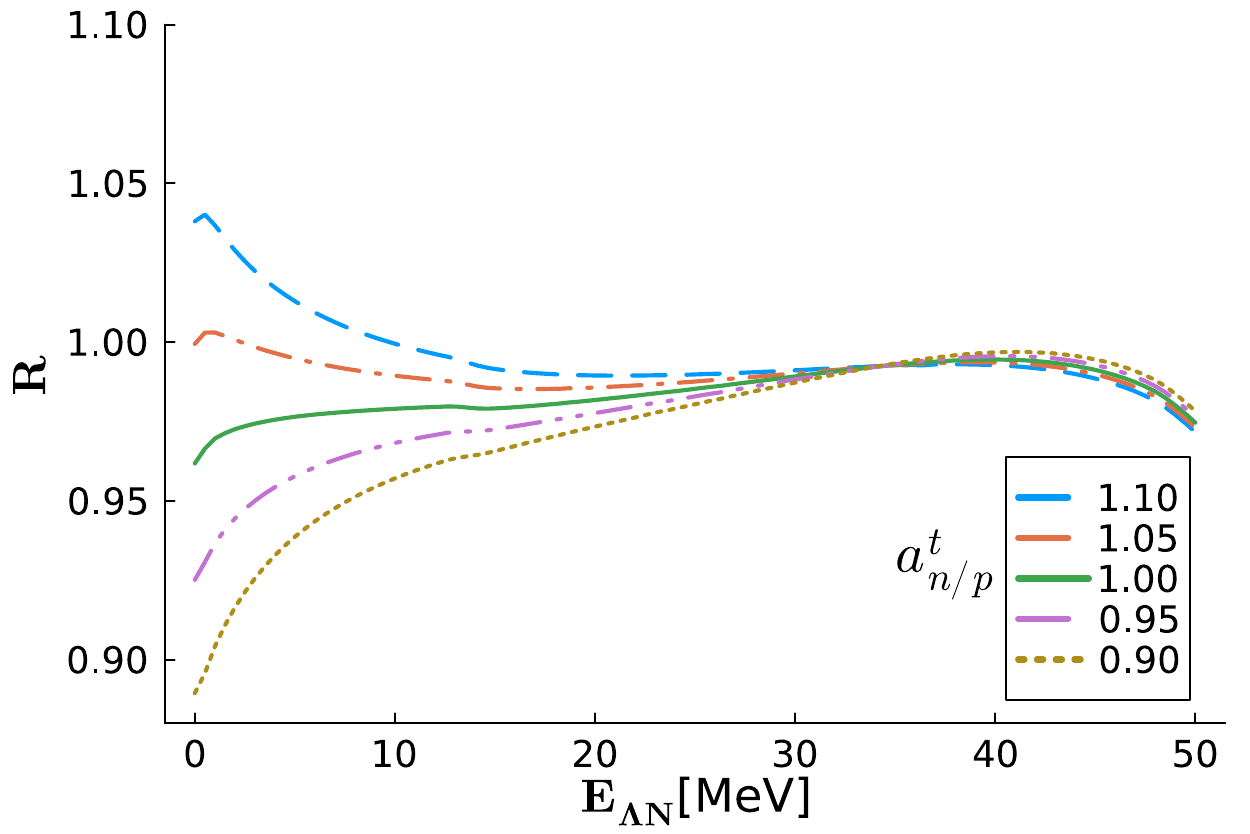}
(a) $0.9\le\cos\theta\le1.0$
\end{subfigure}
\begin{subfigure}{0.48\textwidth}
\includegraphics[width=\textwidth]{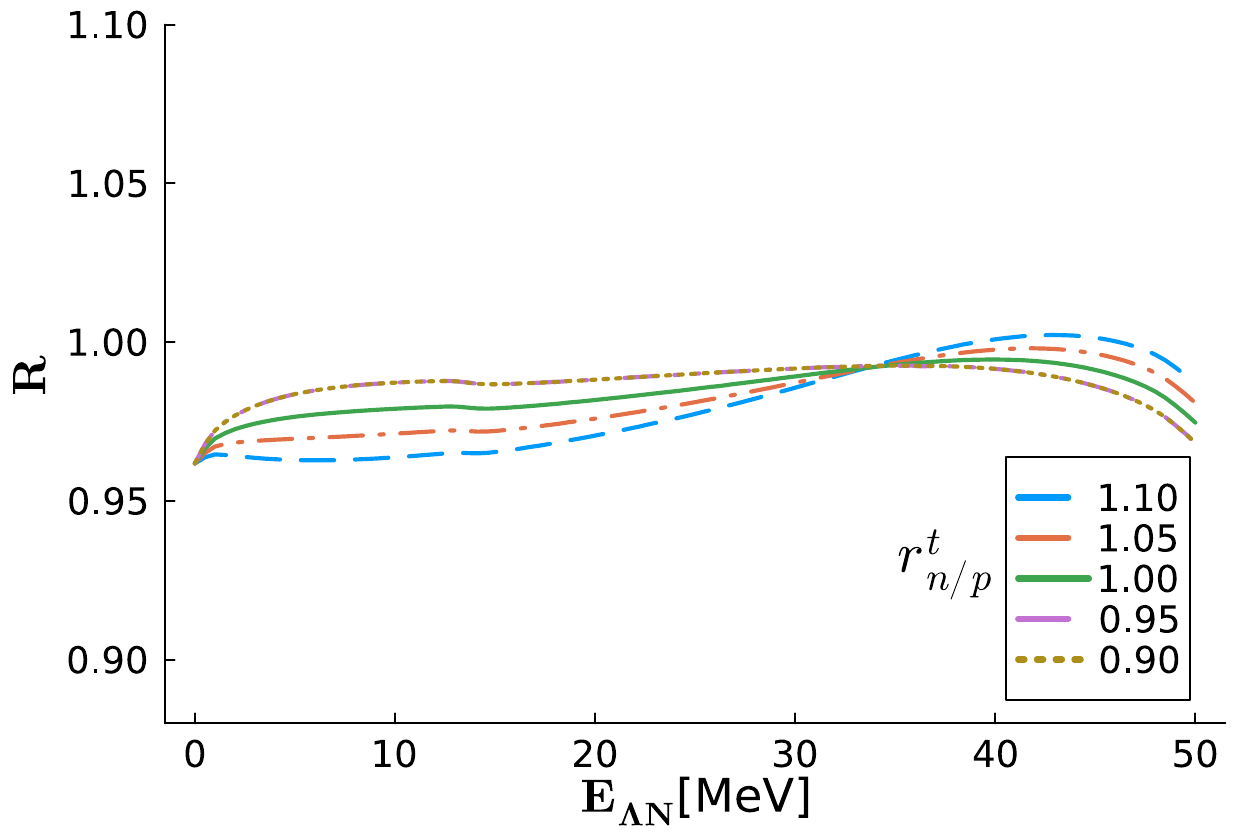}
(b) $0.9\le\cos\theta\le1.0$
\end{subfigure}
\begin{subfigure}{0.48\textwidth}
\includegraphics[width=\textwidth]{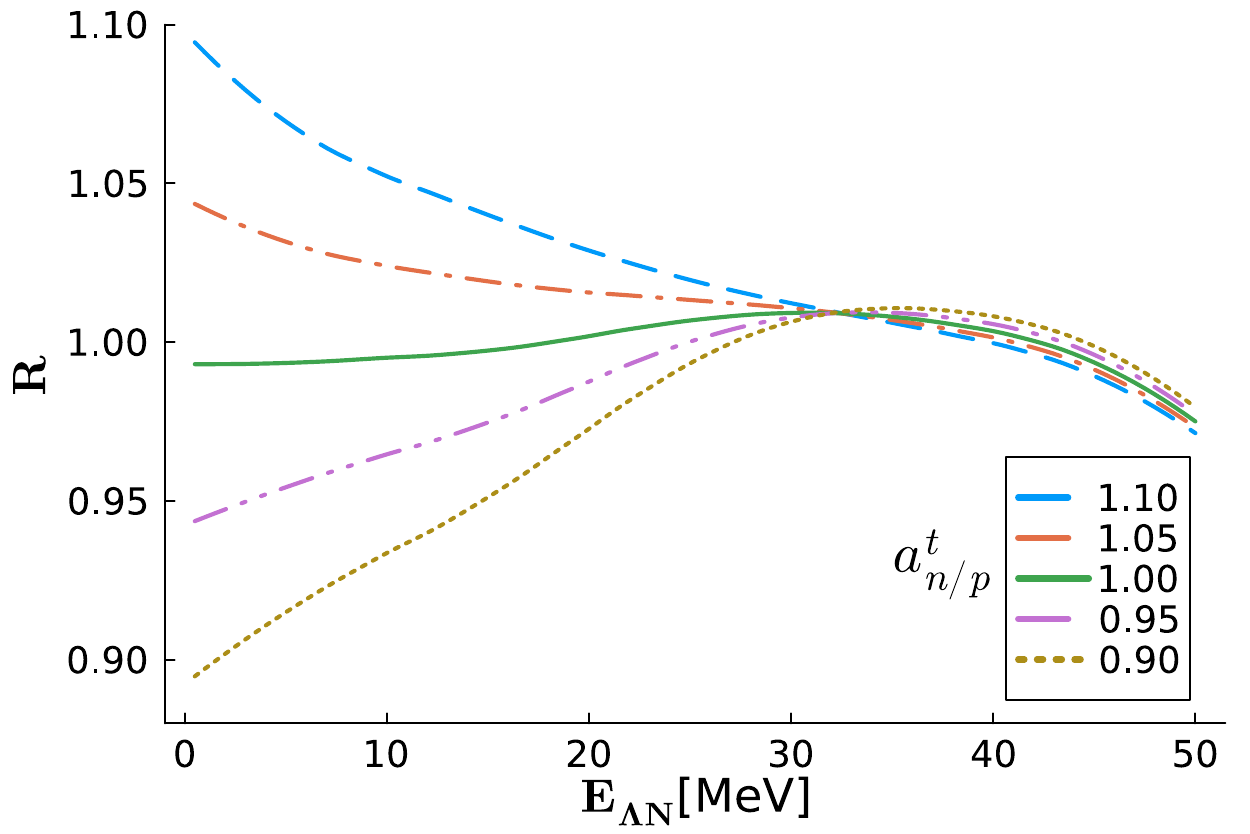}
(c) $0.9\le\cos\theta\le1.0$ and $p_\text{cut} = 150$ MeV/c
\end{subfigure}
\begin{subfigure}{0.48\textwidth}
\includegraphics[width=\textwidth]{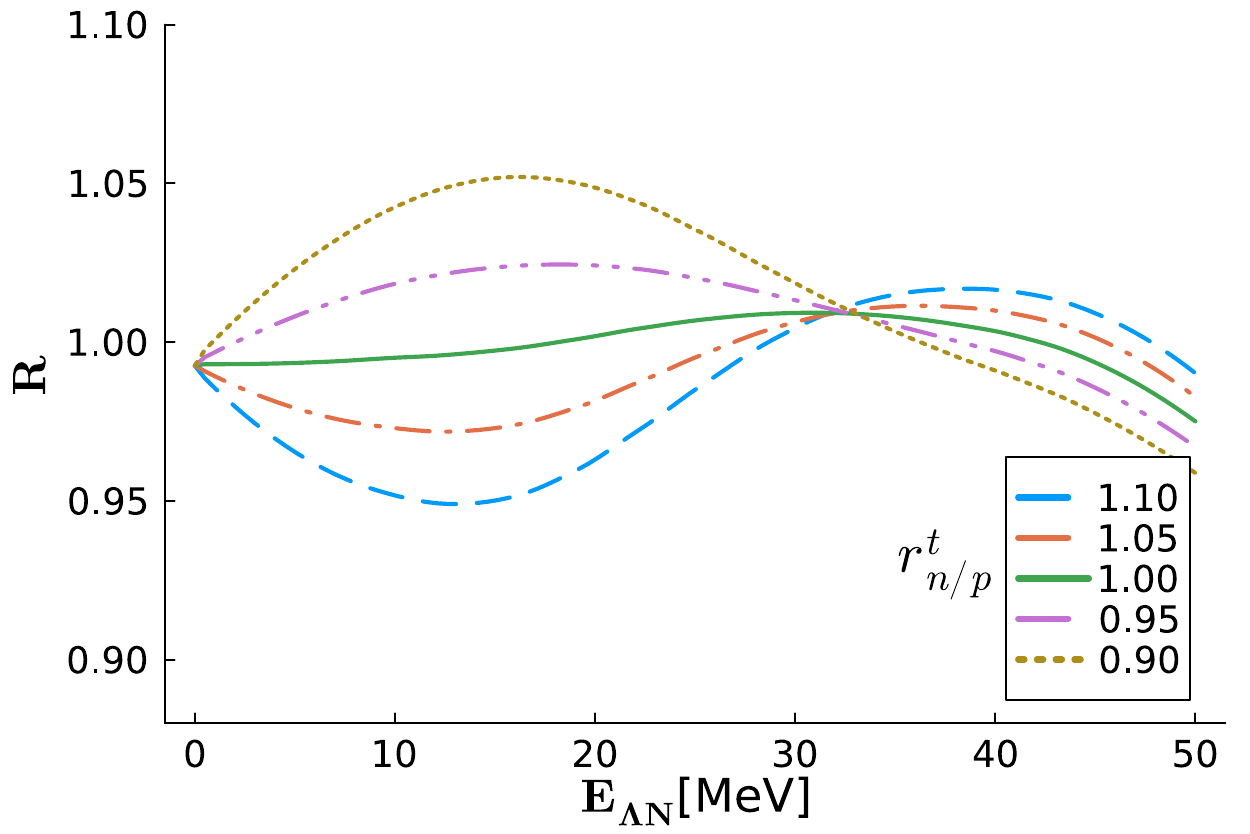}
(c) $0.9\le\cos\theta\le1.0$ and $p_\text{cut} = 150$ MeV/c  
\end{subfigure}
\end{center}
\caption{Ratios of the $\Lambda N$ invariant mass spectra $R$, with two constraint conditions and several values of $a_{n/p}$ and $r_{n/p}$. (a) $0.9\le\cos\theta,1.0$ and values of $a_{n/p}^t=[1.10, 1.05, 1.00, 0.95, 0.90]$. (b) $0.9\le\cos\theta\le1.0$ and several values of $r_{n/p}^t$. (c) $0.9\le\cos\theta\le1.0$ and $p_\text{cut}=150$MeV/c, and several values of $a_{n/p}^t$. (d) $0.9\le\cos\theta\le1.0$ and $p_\text{cut}=150$MeV/c, and several values of $r_{n/p}^t$. The lines [dash, dash-dotted, solid, dash-dot-dotted and dotted] represent the values of $a_{n/p}^t$ or $r_{n/p}^t$ in the order [1.10, 1.05, 1.00, 0.95, 0.90].
}.\label{fig:ar}
\end{figure*}

As in the previous sections, we employ the angular and momentum constraints for reducing background processes. Figure \ref{fig:ar} shows the numerical results of the ratios of $\Lambda N$ invariant mass spectra, $R$ with two selection conditions, only $0.9\le\cos\theta\le1.0$ and both of $0.9\le\cos\theta\le1.0$ and $|\vec{p}_{N}|=150$MeV/c, and several values of $a^t_{n/p}$ and $r^t_{n/p}$. (a) in Fig.\ref{fig:ar} is the ratio with the selection condition $0.9\le\cos\theta\le1.0$, and values of $a_{n/p}^t=[1.10, 1.05, 1.00, 0.95, 0.90]$. (b) is the ratio with the selection condition $0.9\le\cos\theta\le1.0$, and values of $r_{n/p}^t=[1.10, 1.05, 1.00, 0.95, 0.90]$. (c) is the ratio with the selection condition $0.9\le\cos\theta\le1.0$ and $p_\text{cut}=150$MeV/c, and values of $a_{n/p}^t=[1.10, 1.05, 1.00, 0.95, 0.90]$. (d) is the ratio with the selection condition $0.9\le\cos\theta\le1.0$ and $p_\text{cut}=150$MeV/c, and values of $r_{n/p}^t=[1.10, 1.05, 1.00, 0.95, 0.90]$. First, we focus on the deviations between two selection conditions. (a) and (b) are the results with only the angular constraint $0.9\le\cos\theta\le1.0$, and (c) and (d) are the results with both of the constraints $0.9\le\cos\theta\le1.0$ and $p_\text{cut}=150$MeV/c. From these plots, one can say that the momentum selection of the nucleon is better for the sensitivity of the ratio. When comparing (a) with (c), and (b) with (d), their sensitivity to $a^t_{n/p}$ or $r^t_{n/p}$ become higher. We find that deviations up to about 10 percent are observed at the threshold in (b), while in (d), deviations up to around 5 percent are observed around $E_{\Lambda N}=15$ MeV. Next, focus on the behaviors of $a^t_{n/p}$ and $r^t_{n/p}$, respectively. (a) and (c) show that when the value of $a_{n/p}^t$ is large, the value of $R$ becomes large. This result is understood by the definition of the scattering length which represents scattering amplitude at zero momentum. If $a_{n/p}^t$ is larger than 1 at the threshold, $R$ becomes larger than the result with $a_{n/p}^t=1$. (b) and (d) show that effective range $r^t_{n/p}$ changes $R$ oppositely. This is due to the relative sign of the effective range to the scattering length. From Eq.~(\ref{equ:eff_range_t}) and the values of the scattering length and the effective range, $-1/a^t_{YN}$ and $\frac{1}{2}r^t_{YN}p^{*2}$ in Eq.~(\ref{equ:eff_range_t}) have the same sign. Thus the deviations of $a_{n/p}^t$ and $r^t_{n/p}$ cause opposite changes of $R$. Lastly, we mention about nodes around 30MeV in Fig.~\ref{fig:ar}. As seen in these nodes are due to the influence of meson exchange processes being greater than that of the foreground process in this region as seen in Figs.~\ref{fig:ang_lim} and \ref{fig:p_moment}.

%% file: discussion.tex
\section{discussion}\label{sec:dis}
\subsection{Comparison with an old experiment}\label{sec:sigmaN}
\begin{figure}[tb]
\centering
\includegraphics[width=0.48\textwidth]{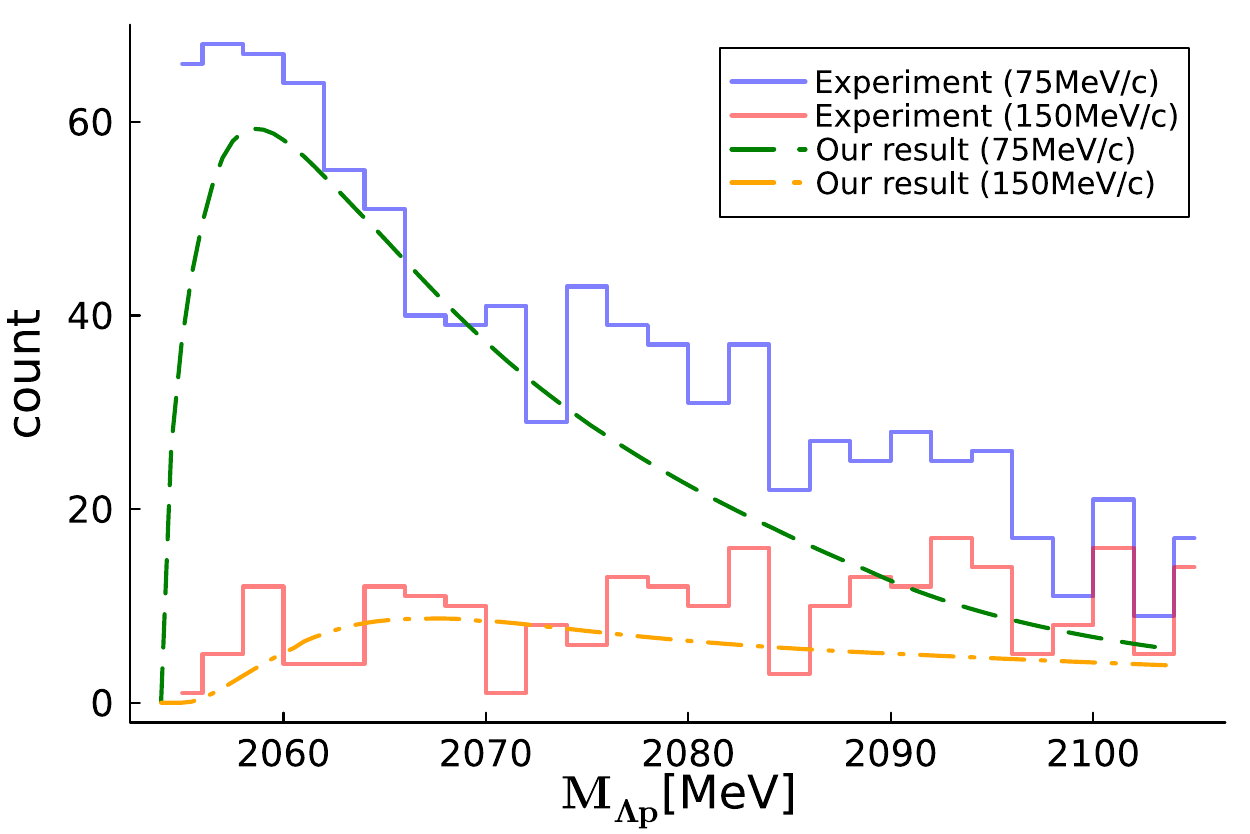}
\caption{A comparison with experimental data. Discrete lines represent experimental data, the line exhibiting higher values represents $p_\text{cut}=$ 75 MeV/c, while the line with lower values represents $p_\text{cut}=$ 150 MeV/c. The dashed line indicates our result with $p_\text{cut} = 75$ MeV/c, while the dash-dotted line corresponds to $p_\text{cut} = 150$ MeV/c.} \label{fig:exp}
\end{figure}

This section compares the calculated $\Lambda p$ invariant mass spectrum for the $K^-d\rightarrow\pi^-\Lambda p$ reaction with experimental data. The plotting area of the $\Lambda p$ invariant mass is the same as in Sec.~\ref{sec:res}, and the contribution of $\Sigma$ exchange is negligibly small. Therefore, the scattering length $a_{\Sigma N}$, which governs the contribution of $\Sigma$ exchange, does not make a difference regardless of its value. Figure~\ref{fig:exp} depicts comparison of the computed results with experimental data \cite{braun_1977}. All lines in Fig.~\ref{fig:exp} satisfy the common selection, $0.9\le\cos\theta\le1.0$ in the c.m. frame. The histograms represent experimental data with incident momenta ranging from 680 to 840 MeV/c.
The histogram exhibiting higher values shows the data with $p_\text{cut}=$ 75 MeV/c which is the lower bound of the momentum of selected protons, while the lower histogram represents the data with $p_\text{cut}=$ 150 MeV/c.
Our computations are conducted with 760 MeV/c incident momentum. The dashed line indicates our result with $p_\text{cut} = 75$ MeV/c, while the dash-dotted line corresponds to $p_\text{cut} = 150$ MeV/c. The y-axis is represented in count numbers from experimental data. Scaling of our results is performed considering the experimental condition which is an incident flux of 7.1 $\mu$b. In the numerical results, for $p_\text{cut}=75$ MeV, a gradual peak is seen around 2065 MeV, while for $p_\text{cut}=150$ MeV, a peak is seen around 2060 MeV, followed by a decrease, and two results approximately converge around 2100 MeV. These numerical computation results are quantitatively consistent with the experimental data. This indicates that the selection of final-state kinematics contributes the good quantitative agreement between theory and experiment.

\subsection{Correction of Coulomb interaction}
\begin{figure}[tb]
\centering
\includegraphics[width=0.48\textwidth]{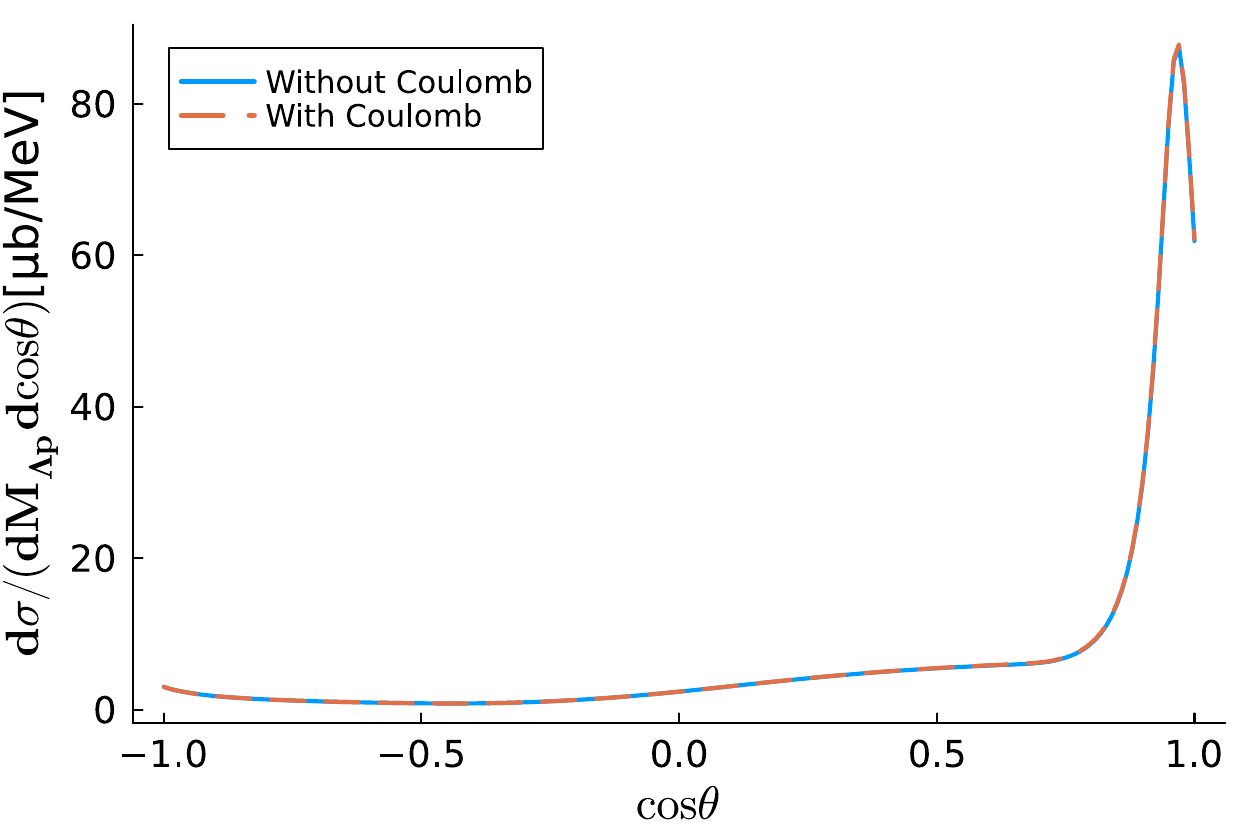}
\caption{Angular dependence of the differential cross section for the $K^-d\rightarrow\pi^-\Lambda p$ reaction at an excitation energy $E_{\Lambda p}=10$ MeV. Solid lines represent the result without considering Coulomb interaction, while dashed lines represent that considering Coulomb interaction.}\label{fig:Coulomb}
\end{figure}

It may be a question that the inclusion of the Coulomb interaction affect our numerical outcomes. To explore this, we examine the case where the Coulomb interaction influences the second meson-baryon interaction. For the $\pi^-\Lambda p$ final state, $\pi^-$ exchange process (e1) in Fig.~\ref{fig:p_process} has a $\pi^-p\rightarrow\pi^-p$ interaction in the process of the second scattering. Since each particle existing simultaneously has a charge of opposite sign, Coulomb interaction is accounted for the amplitude. The interaction can be incorporated into the strong interaction amplitude which takes into consideration the effects of both the electromagnetic and strong interactions on the scattering process. This formulation is written in \cite{hashimoto_1984,aoki_2017}. Up to the p-wave partial wave analysis~\eqref{equ:partial}, the spin-independent part and the spin-dependent part of the meson-baryon amplitude, denoted as $F$ and $G$ respectively, are written by
\begin{align}
F &= T^0 + (2T^{1+} + T^{1-}) e^{2i\phi_1}\cos\theta - 8\pi W f_C,\\
G &= (T^{1+} - T^{1-})e^{2i\phi_1},
\end{align}
where Coulomb amplitude $f_C$ and Coulomb phase shift $\phi_1$ are described as
\begin{align}
f_c &= - \frac{\alpha}{2k\nu\sin^2(\theta/2)} \exp[-i\frac{\alpha}{\nu}\log(\sin^2\frac{\theta}{2})]\\
\phi_1 &= \tan^{-1}\frac{\alpha}{n\nu}.\\
\intertext{Relative velocity between meson and baryon $\nu$ is calculated by}
\nu &= \frac{p(e + E)}{eE},
\end{align}
where $p$, $e$, and $E$ are meson momentum, meson energy, and baryon energy, respectively. The Coulomb interaction is parametrized by $\alpha$, fine-structure constant. Using these formulae, the
numerical result of angular dependence for both cases, without and with Coulomb interaction, are depicted as shown in Fig.~\ref{fig:Coulomb}. This figure shows that the Coulomb interaction in the intermediate state of the $\pi$ exchange process has little contribution to the outcomes. While two-body Coulomb interaction strongly affects the forward direction, the forward Coulomb interaction in the intermediate state is averaged out by integration, resulting in little impact. From this study, it can be concluded that there is little need to consider the Coulomb interaction.

%% file: summary.tex
 \section{Summary}\label{sec:sum}
In this article, we have calculated the $\Lambda N$ invariant mass spectra of the $K^-d\rightarrow\pi\Lambda N$ reaction with in-flight kaons and studied how isopin symmetry breaking in the $\Lambda N$ system affects the ratio of the spectra between $K^- d\rightarrow \pi^-\Lambda p$ and $K^- d\rightarrow \pi^0\Lambda n$ reactions by utilizing $\Lambda N$ final state interaction. With regarding the momentum effect, p-wave components for meson-baryon amplitude and spin dependent terms for hyperon-baryon system have been considered. We have employed two constraints, kinematics where the direction of outgoing pions is forward and the momentum of outgoing nucleons is higher enough. These selections have served as reductions of background processes.  Calculations have been conducted at around 1000MeV/c of incident kaon's momentum, and the results obtained are quantitatively consistent with a past experimental data in the vicinity of 680-840 MeV/c.

The degree of isospin symmetry breaking in the $\Lambda N$ system can be deduced from that in the spectrum of the $K^-d\rightarrow\pi\Lambda N$ reaction. Through appropriate kinematic selections, we have yielded results where 10 percent difference in the scattering length ratio for the spin-triplet state $a^t_{n/p}=a^t_n/a^t_p$ corresponds to approximately 10 percent difference in the spectrum ratio $R=2\sigma_{\pi^0\Lambda n}/\sigma_{\pi^-\Lambda p}$, reflecting each other to a similar extent. Similarly, an analysis of the apin-triplet state in the effective range has revealed the difference for $R$ of up to 5 percent at most.

%% file: appendix.tex
\begin{appendix}\label{sec:app}
\section{Linear combinations of hyperon-nucleon amplitudes}\label{sec:linear}
Hyperon-nucleon amplitude $T_{YN}$ is divided into two parts in the spin space, identity matrix part $\bm{1}$ and spin-spin interaction part $\vec{\sigma}_Y\cdot\vec{\sigma}_N$. There are two kinds of spin states, singlet (total spin 0) and triplet (total spin 1). These scattering amplitudes are calculated by using the spin singlet state $\ket{s}$ and the spin triplet state $\ket{t}$ as follows:
\begin{align}
T^s_{YN} &= T^1_{YN}\bra{s}\bm{1}\ket{s} + T^\sigma_{YN}\bra{s}\vec{\sigma}_Y\cdot\vec{\sigma}_N\ket{s},\\
T^t_{YN} &= T^1_{YN}\bra{t}\bm{1}\ket{t} + T^\sigma_{YN}\bra{t}\vec{\sigma}_Y\cdot\vec{\sigma}_N\ket{t},
\end{align}
Inner product of Pauli matrices can be described by quadratic Casimir operator as $\vec{\sigma}_Y\cdot\vec{\sigma}_N = \frac{1}{2}[(\vec{\sigma}_Y +\vec{\sigma}_N)^2 - \vec{\sigma}_Y^2 - \vec{\sigma}_N^2]$ and spin operator is half of the Pauli matrices as $\vec{s} = \vec{\sigma}/2$, thus each amplitude is represented as follows:
\begin{align}
T^s_{YN} &= T^1_{YN} - 3T^\sigma_{YN}, \\
T^t_{YN} &= T^1_{YN} + T^\sigma_{YN}.
\end{align}
Conversely, expressing $T^1_{YN}$ and $T^\sigma_{YN}$ in terms of $T^s_{YN}$ and $T^t_{YN}$, we obtain Eqs.~\eqref{equ:T1} and \eqref{equ:Tsigma}.

\section{Spin matrix of each amplitude} \label{sec:spin_mat}
Spin matrices of the amplitudes are generally written as the linear combination of the basis made from $S_a^\dag$, $\sigma_i$, and $(\sigma_i)^T$ as
\begin{equation}
\mathcal{T}_a = AS_a^\dag + B_iiS_a^\dag(\sigma_i)^T + C_ii\sigma_iS_a^\dag + D_{ij}\sigma_iS_a^\dag(\sigma_j)^T,\\
\end{equation}
where $A$, $B_i$, $C_i$, and $D_{ij}$ are coefficients, and their matrix elements for the $\Lambda$ and $N$ spinors, $\chi_\Lambda$ and $\chi_N$, are given as $(\chi_\Lambda)^T \mathcal{T}_a \chi_N$.

Taking the trace over these bases yields the following rtesult:
\begin{align}
&\frac{1}{3}\tr[\mathcal{T}^{(\alpha)}_a\mathcal{T}^{(\beta)}_a]\notag\\
=&\ A^{(\alpha)}A^{*(\beta)} + B^{(\alpha)}_iB^{*(\beta)}_i + C^{(\alpha)}_iC^{*(\beta)}_i + D^{(\alpha)}_{ij}D^{*(\beta)}_{ij}\notag\\
&+\frac{1}{3}(A^{(\alpha)}D^{*(\beta)}_{ii}+D^{(\alpha)}_{ii}A^{*(\beta)})\notag\\
&+\frac{1}{3}(B^{(\alpha)}_iC^{*(\beta)}_i+C^{(\alpha)}_iB^{*(\beta)}_i)\notag\\
&+\frac{1}{3}\varepsilon_{ijk}(-B_i^{(\alpha)}D^{*(\beta)}_{jk}-D^{(\alpha)}_{ij}B^{*(\beta)}_k)\notag\\
&+\frac{1}{3}\varepsilon_{ijk}(-B_i^{(\alpha)}D^{*(\beta)}_{jk}-D^{(\alpha)}_{ij}B^{*(\beta)}_k)\notag\\
&+\frac{1}{3}(D^{(\alpha)}_{ij}D^{*(\beta)}_{ji}-D^{(\alpha)}_{ii}B^{*(\beta)}_{jj}).\notag\\
\end{align}
The following subsections write down the coefficients for each process.

\subsection{Impulse process}\label{subsec:impulse}
We divide spin matrices of meson-baryon amplitude $T_{ij}$ into spin independent part $F_{ij}$ and spin dependent part $G_{ij}$:
\begin{equation}
T_{ij}(W,\theta) = F_{ij}(W,\theta)-iG_{ij}(W,\theta)~\vec{\sigma}\cdot(\hat{k}_j\times\hat{k}_i),\label{equ:mes-bar}
\end{equation}
where $\hat{k}_i$ and $\hat{k}_j$ are unit vectors along the momentum of the mesons in each channel $i$ and $j$, respectively. The coefficients of amplitude for impulse process~\eqref{equ:imp} is calculated as follows:
\begin{equation}
\begin{aligned}
A^{(I)}&=F_{K^-N\rightarrow\pi\Lambda}\tilde{\varphi}(|\vec{p}_N|),\\
B^{(I)}_i&=0,\\
C^{(I)}_i&=-G_{K^-N\rightarrow\pi\Lambda}\sin\theta_i\tilde{\varphi}(|\vec{p}_N|),\\
D^{(I)}_{ij}&=0.
\end{aligned}
\end{equation}

\subsection{Hyperon exchange process}\label{subsec:hyperon}
Using Eq.~\eqref{equ:mes-bar}, with denoting $F_{K^-N\rightarrow\pi Y}$ and $G_{K^-N\rightarrow\pi Y}$ as $F$ and $G$, respectively, hyperon scattering amplitude~\eqref{equ:hypex} can be expressed as 
\begin{equation}
\begin{aligned}
\mathcal{T}_a^{(Y)}&= (\mathcal{T}^1_a + \mathcal{T}^\sigma_a)\int \frac{d^3\vec{q}}{(2\pi)^3} \frac{2M\ \tilde{\varphi}(|\vec{q} + \vec{Q}|)}{(q_0)^2 - |\vec{q}|^2 - M^2 + i\epsilon},\\
\mathcal{T}^1_a &=  T^1_{YN} \left(F S^{\dag}_a - iG \sin\theta_i \sigma^i S^\dag_a \right),\\
\mathcal{T}^\sigma_a &=  T^\sigma_{YN} \left(F S^{\dag}_a - iG\sin_i(2S^\dag_a(\sigma_i)^T - \sigma_iS^\dag_a)\right),
\end{aligned}
\end{equation}
where the angle is defined as $\sin\theta_i = \left( \hat{p}_\pi \times \hat{p}_{\bar{K}} \right)_i$. Therefore, the coefficients for hyperon exchange processes are calculated as follows:
\begin{equation}\label{equ:spin_sel}
\begin{aligned}
A^{(Y)} &= 2M_YT^t_{YN}F\Phi_0,\\
B^{(Y)}_i &= 2M_Y\left(-\frac{1}{2}T^t_{YN} - \frac{1}{2}T^s_{YN}\right)G\Phi_0\sin\theta_i,\\
C^{(Y)}_i &= 2M_Y\left(-\frac{1}{2}T^t_{YN} + \frac{1}{2}T^s_{YN}\right)G\Phi_0\sin\theta_i,\\
D^{(Y)}_{ij} &= 0,
\end{aligned}
\end{equation}
where $\Phi_0=\Phi_0(Q, q_0; M)$ ($M$ denotes the mass of propagating particle) is integral with the deuteron wave function defined and calculated in Appendix~\ref{sec:integral}. Equation~\eqref{equ:spin_sel} demonstrates the spin selection rule, where the pion forward part of the scattering ($\sin\theta_i \sim 0$, expressed in another way, $\hat{p}_\pi /\!/ \hat{p}_K$) is dominant for the spin triplet state.

\subsection{Meson exchange process}
Amplitudes for meson exchange processes are expressed by Eqs.~\eqref{equ:kex} and \eqref{equ:piex}, their general form is expressed as
\begin{align}
\mathcal{T}^{(\pi,\bar{K})}_a = \int&\frac{d^3\vec{q}}{(2\pi)^3}\ T_2(W_2,\hat{q}\cdot\hat{p}_2)\ S_a^\dag\ [T_1(W_1,\hat{p}_1\cdot\hat{q})]^T\notag\\
&\times\frac{\tilde{\varphi}(|\vec{q}+\vec{Q}|)}{(q_0)^2 -|\vec{q}|^2 - M^2 + i\epsilon},
\end{align}
where $T_1$ and $T_2$ represent the scattering amplitudes for the first and second step scatterings, respectively. Using the decomposition into $F$ and $G$:
\begin{equation}
\begin{aligned}
T_1 &= F_1 -iG_1\ \vec{\sigma}\cdot(\hat{q}\times\hat{p}_{K^-}),\\
T_2 &= F_2 -iG_2\ \vec{\sigma}\cdot(\hat{p}_\pi\times\hat{q}),
\end{aligned}
\end{equation}
and with the partial wave decomposition up to p-wave as
\begin{equation}
\begin{aligned}
F(W,\theta) &= F^{(0)}(W) + F^{(1)}(W)P_1(\theta),\\
G(W,\theta) &= G^{(1)}(W),
\end{aligned}  
\end{equation}
the coefficients for the meson exchange process are as follows:
\begin{widetext} 
\begin{equation*}
\begin{aligned}
A^{(\pi,\bar{K})}=&\ F_2^{(0)}F_1^{(0)}\Phi_0 \\
&+ F_2^{(0)}F_1^{(1)}(\hat{k}_1\cdot\hat{Q})\Phi_1 + F_2^{(1)}F_1^{(0)}(\hat{Q}\cdot\hat{k}_2)\Phi_1\\
&+ F_2^{(1)}F_1^{(1)} \bigg[\frac{1}{3}(\hat{k}_2\cdot\hat{k}_1)(\Phi_0-\Phi_2)+(\hat{Q}\cdot\hat{k}_2)(\hat{p}_1\cdot\hat{Q})\Phi_2\bigg],\\
B^{(\pi,\bar{K})}_i=&\ F_2^{(0)}G_1^{(1)}(\hat{Q}\times\hat{k}_1)_i\Phi_1\\
&+ F_2^{(1)}G_1^{(1)}\bigg[\frac{1}{3}(\hat{p}_2\times\hat{p}_1)_i(\Phi_0-\Phi_2)+(\hat{p}_2\cdot\hat{Q})(\hat{Q}\times\hat{p}_1)_i\Phi_2\bigg],\\
\end{aligned}
\end{equation*}
\begin{equation}
\begin{aligned}
C^{(\pi,\bar{K})}_i=&\ G_2^{(1)}F_1^{(0)}(\hat{k}_2\times\hat{Q})I_1 \\
&+ G_2^{(1)}F_1^{(0)}\bigg[\frac{1}{3}(\hat{p}_2\times\hat{p}_1)_i(\Phi_0-\Phi_2)+(\hat{p}_2\times\hat{Q})_i(\hat{Q}\cdot\hat{p}_1)\Phi_2\bigg],\\
D^{(\pi,\bar{K})}_{ij}=&\ G_2^{(1)}G_1^{(1)}\bigg[\frac{1}{3}\big((\hat{p}_1)_i(\hat{p}_2)_j - (\hat{p}_1\cdot\hat{p}_2)\delta_{ij}\big)(\Phi_0-\Phi_2)+(\hat{p}_2\times\hat{Q})_i(\hat{Q}\times\hat{p}_1)_j\Phi_2\bigg],
\end{aligned}
\end{equation}
\end{widetext}
where $\Phi_n=\Phi_n(Q, q_0; M)$ (here, $n=0,1,2$) are integrals with the deuteron wave function defined and calculated in the next section.

\section{Integral with the deuteron wave function}\label{sec:integral}
Integral with the deuteron wave function $\Phi_n(Q,q_0;M)$ is defined as
\begin{equation}
\Phi_n(Q,q_0;M) = \int\frac{d^3\vec{q}}{(2\pi)^3} \frac{P_n(\hat{q}\cdot\hat{Q})}{q_0^2 - |\vec{q}|^2 - M^2 + i\epsilon}\tilde{\varphi}(|\vec{q}+\vec{Q}|),
\end{equation}
where $P_n$ is a Legendre polynomial. Rewriting $\mu = \sqrt{M^2 - q_0^2}$ and using deuteron wave function (\ref{equ:deuteron}), $\Phi_n(Q,\mu)$ is parametrized as
\begin{equation}
\Phi_n(Q,\mu) = N\sum_{i=1}^{11} C_i I_n(Q,\mu;m_i),
\end{equation}
where $N$ and $C_i$ represent normalization constants given in the deuteron wave function. The integral $I_n(Q,\mu;m)$ is defined as
\begin{equation}
I_n(Q,\mu;m) = - \int\frac{d^3\vec{q}}{(2\pi)^3} \frac{P_n(\hat{q}\cdot\hat{Q})}{|\vec{q}|^2 + \mu^2 + i\epsilon}\frac{1}{(\vec{q}+\vec{Q})^2 +m^2}.  \label{equ:loop}
\end{equation}
By performing a Fourier transformation:
\begin{equation}
\frac{1}{(\vec{q}+\vec{Q})^2 + m^2} = \int d^3\vec{r}\  \frac{e^{-m|\vec{r}|}}{4\pi |\vec{r}|}e^{-i(\vec{q}+\vec{Q})\cdot\vec{r}},
\end{equation}
and utilizing spherical harmonics $Y_l^m$ and spherical Bessel functions $j_l$ as:
\begin{align}
&e^{-i\vec{q}\cdot\vec{r}}=4\pi\sum_{l=0}^\infty\sum_{m=-l}^{+l}(-i)^lj_l(qr)Y_l^{m*}(\hat{q})Y_l^m(\hat{r}),\\
&P_n(\hat{q}\cdot\hat{Q})=\frac{4\pi}{2n+1}\sum_{k=-n}^{+n}Y_n^{k*}(\hat{q})Y_n^k(\hat{Q}),
\end{align}
$I_n$ can be expressed as a double integral involving Bessel functions $J_l$:
\begin{equation}
\begin{aligned}
I_n(Q,\mu;m) &= \frac{(-1)^{n+1}}{4\pi\sqrt{Q}}\int_0^\infty dq \Bigg[ \ \frac{q^{\frac{3}{2}}}{q^2+\mu^2+i\epsilon} \\
& \ \ \ \times\int_0^\infty dr e^{-mr} J_{n+\frac{1}{2}}(Qr)J_{n+\frac{1}{2}}(qr)\Bigg].\label{equ:doublebessel}
\end{aligned}
\end{equation}
The integral over $r$ in Eq.~\eqref{equ:doublebessel}, Laplace transformation of two Bessel functions' product yields:
\begin{align}
&\int_0^\infty dr \ e^{-mr} J_{n+\frac{1}{2}}(Qr) J_{n+\frac{1}{2}}(qr) \notag\\
=&\frac{1}{\pi\sqrt{Qq}} Q_n\left(\frac{m^2+Q^2+q^2}{2Qq}\right),
\end{align}
where $Q_n$ is Legendre function of the second kind. The specific expression for $Q_n(z)$ up to $n=2$ are as follows:
\begin{align}
Q_0(z) &= \frac{1}{2}\ln(\frac{z+1}{z-1}),\\
Q_1(z) &= \frac{1}{2}z\ln(\frac{z+1}{z-1})-1,\\
Q_2(z) &= \frac{1}{4}(3z^2-1)\ln(\frac{z+1}{z-1})-\frac{3}{2}z.
\end{align}
Thus, the integral $I_n$ can be expressed as
\begin{equation}
I_n(Q,\mu;m) = \frac{(-1)^{n+1}}{4\pi^2Q}\int_0^\infty dq\frac{qQ_n\left(\frac{m^2+Q^2+q^2}{2Qq}\right)}{q^2+\mu^2+i\epsilon}.
\end{equation}
For the case of $n=0$ and $2$, the analytical expressions for $I_n$ are obtained as follows:
\begin{widetext}
\begin{equation}
I_0(Q,\mu;m)= \frac{i}{8\pi Q}\ln(\frac{m+\mu+Qi}{m+\mu-Qi}),
\end{equation}
\begin{equation}
\begin{aligned}
I_2(Q,\mu;m)= &\frac{3}{32\pi\mu Q^2}\left(m^2+\mu m-\mu^2+Q^2\right) + \frac{3i}{64\pi\mu^2Q^3}(m^2+Q^2)^2\ln\left(\frac{m+Qi}{m-Qi}\right)\\
                &-\frac{i}{64\pi\mu^2Q^3}\left[3Q^4+2(3m^2-\mu^2)Q^2 + 3(m^2 - \mu^2)^2\right]\ln\left(\frac{m+\mu+Qi}{m+\mu-Qi}\right).
\end{aligned}
\end{equation}
\end{widetext}
\end{appendix}